\documentclass[twocolumn,showpacs,preprintnumbers,amsmath,amssymb]{revtex4-1}

\usepackage{graphicx}
\usepackage{dcolumn}
\usepackage{bm}

\begin{document}


\title{Electronic and magnetic properties of 
 the 2H-NbS$_2$ intercalated by 3d transition metal atoms}
\author{S.\ Mankovsky}
\affiliation{%
  Department  Chemie,  Physikalische  Chemie,  Universit\"at  M\"unchen,
  Butenandstr.\  5-13, 81377 M\"unchen, Germany\\}
\author{S.\ Polesya} 
\affiliation{%
  Department  Chemie,  Physikalische  Chemie,  Universit\"at  M\"unchen,
  Butenandstr.\ 5-13, 81377  M\"unchen, Germany\\}
\author{H.\ Ebert} 
\affiliation{%
  Department  Chemie,  Physikalische  Chemie,  Universit\"at  M\"unchen,
  Butenandstr.\ 5-13, 81377  M\"unchen, Germany\\}
\author{W.\ Bensch} 
\affiliation{%
Inst.\ f\"ur Anorgan.\ Chemie,  Universit\"at Kiel,
Olshausenstr.\ 40,  24098 Kiel, Germany\\}

\date{\today}
             
\begin{abstract}
The electronic structure and magnetic properties of the 2H-NbS$_2$
compound intercalated by Cr, Mn and Fe, have been investigated by means
of the Korringa-Kohn-Rostoker (KKR) method. The calculations demonstrate
easy plane magneto-crystalline anisotropy (MCA) of Cr$_{1/3}$NbS$_2$
monotonously decreasing towards the Curie temperature in line with the
experimental results. The modification of the electronic structure
results in a change of the easy axis from in-plane to out-of-plane. It
is shown, that for Cr$_{1/3}$NbS$_2$ and  Mn$_{1/3}$NbS$_2$ the in-plane MCA and
Dzyaloshinskii-Moriya interactions results in a helimagnetic
structure along the hexagonal $c$ axis, following the experimental
observations. The negative exchange interactions in 
the Fe$_{1/3}$NbS$_2$ compound results in a non-collinear frustrated
magnetic structure if the MCA is not taken into account. It is shown,
however, that a strong MCA along the hexagonal $c$ axis leads to a
magnetic ordering referred to as an ordering of the third kind, which
was observed experimentally. 
\end{abstract}

\pacs{Valid PACS appear here}
\maketitle

\section{Introduction}

The transition metal dichalcogenides (TMDC) are in the focus of extensive  
investigations for several decades. Although most of TMDC systems are
nonmagnetic, they allow intercalation of magnetic $3d$ elements, $M$,
(see, e.g. review article \cite{FY87}) leading to the formation 
of materials exhibiting a great variety of magnetic properties dependent on   
the host system, on the type and concentration of elements $M$. Some examples 
for this are the large magnetoresistance (MR) in Fe$_{0.28}$TaS$_2$ single crystal
\cite{HCM+15} and the extremely high magnetic anisotropy and anomalous Hall
effect (AHE) in Fe$_{1/4}$TaS$_2$ \cite{CLM+08,CKA+09} recently
discussed in the literature. 
In the present work we will focus on NbS$_2$ intercalated with Cr, Mn, 
and Fe. For the concentration 1/3 of the 3$d$ element, 
these systems exhibit the trend to the formation of an ordered compound,
called in the literature either $M$Nb$_3$S$_6$ or $M_{1/3}$NbS$_2$,
with an $\sqrt{3} \times \sqrt{3}$ arrangement of 
the $M$ atoms within the layer.
If these compounds are magnetically ordered, they show a non-vanishing
Dzyaloshinskii-Moriya (DM) interaction because of the lack of 
inversion symmetry. This anisotropic exchange interaction is in particular 
responsible for the formation of a helimagnetic (HM) structure under normal conditions. 
However, an applied external magnetic field leads in some materials 
to the formation of a complicated magnetic texture as for example 
skyrmions having nontrivial topological properties \cite{WKL+13,BH94}. 

Experimentally, it was observed that the compound Cr$_{1/3}$NbS$_2$ 
is a metallic ferromagnet, with a Curie temperature $T_C = 163$~K and a Cr
magnetic moment of $2.9~\mu_B$ \cite{HP70}. A similar
magnitude for the magnetic moment was found also by Miyadai et
al.\cite{MKK+83}, while the Curie 
temperature reported is somewhat smaller ($T_C = 127$~K). 
Performing small-angle neutron diffraction measurements, these authors could
give evidence for a long-period helimagnetic structure in
Cr$_{1/3}$NbS$_2$ along the hexagonal $c$-axis, with Cr magnetic moments
lying within plane perpendicular to the $c$-axis. 
These results have been confirmed by other experimental groups 
by using Lorentz microscopy and small-angle electron diffraction
\cite{TKT+12}, as well as muon spin rotation and relaxation measurements 
\cite{BGT+15}. The experimental investigations using 
a superconducting quantum interference device (SQUID) magnetometer
\cite{KNK+09} demonstrated a transition from the commensurate to the
incommensurate helimagnetic structure in the presence of a magnetic field 
directed perpendicular to the hexagonal $c$-axis. A similar bihaviour 
has also been found for Mn$_{1/3}$NbS$_2$, although no neutron 
scattering results are available so far that show
explicitly the helimagnetic structure in Mn$_{1/3}$NbS$_2$.  

The transition from the helimagnetic structure to the state with an 
incommensurate chiral magnetic soliton lattice (CSL) in
Cr$_{1/3}$NbS$_2$ in the presence of a magnetic field $\vec{H} \perp c$ 
 has been recently investigated by several groups \cite{TKT+12,TKN+13}. 
They have found that the size and number of magnetic domains in the CSL
state, having spin orientation along the magnetic field and separated by
$360^o$ domain walls, can be easily manipulated by the 
magnetic field. This leads in particular to interesting
transport properties in the system, such as a noticeable negative
magneto-resistance (MR) along the chiral $c$ axis \cite{WKL+13},
observed in a wide range of temperature below the
incommensurate-commensurate phase transition 
\cite{TKN+13,GMP+13}. Togawa et al. \cite{TKN+15,TMK+16}
demonstrated (using a combination of magneto-transport measurements and
Lorentz transmission electron microscopy) the step-wise and hysteretic
changes of the MR due to soliton confinement in a small
Cr$_{1/3}$NbS$_2$ crystal with the CSL. 
These properties make Cr$_{1/3}$NbS$_2$ attractive
for spintronic applications. 

The Fe$_{1/3}$NbS$_2$ compound, in contrast to Cr$_{1/3}$NbS$_2$ and
Mn$_{1/3}$NbS$_2$, exhibits antiferromagnetic ordering with the N\'eel 
temperature reported to be $T_{N} =
100$~K \cite{LRI71}, $47$~K \cite{GBR+81}, or $44$~K
\cite{YMT+04}. The analysis of  magnetic neutron scattering data
\cite{LRI71} as well as of M\"ossbauer spectra \cite{GBR+81} 
suggest a collinear alignment of the Fe magnetic moments along the
hexagonal $c$-axis. So far, however, there are no theoretical investigations on
the magnetic structure in this material.

In the present work we present a comparison of magnetic properties of
Cr$_{1/3}$NbS$_2$, Mn$_{1/3}$NbS$_2$ and Fe$_{1/3}$NbS$_2$ compounds,
obtained within first principles calculations, demonstrating
the importance of the 
relativistic effects for their magnetic properties and discuss the
differences between the compounds.

\section{Computational details}

\label{SEC:Computational-scheme}

The present investigations on the magnetic properties of NbS$_2$
intercalated with Cr, Mn, and Fe are based on first-principles
electronic structure calculations which have been performed using the fully
relativistic KKR Green function method \cite{SPR-KKR6.3,EKM11}.
The calculations have been done in the
framework of the local spin density approximation  (LSDA) to 
density functional theory (DFT) using the parametrization for
the exchange and correlation potential as given by
Vosko {\em et al.} \cite{VWN80}. For the angular momentum
expansion of the Green function a cutoff of
$l_{max} = 3$ was applied.  

The magneto-crystalline anisotropy (MCA) at $T = 0$~K
was investigated via first-principles calculations of the magnetic
torque $\vec{T}_i^{(\hat{e}_i)}=-\partial E(\{\hat{e}_k\})/\partial
\hat{e}_i \times \hat{e}_i$ acting on the magnetic moment $\vec{m}_i =
m_i\hat{e}_i$ at atomic site $i$ and oriented along the direction of the
total magnetization $\vec{M}$. The torque projection onto the direction
$\hat{u}$, $T_{\hat{u}} (\theta, \phi) = \vec{T}_i^{(\hat{e}_i)}\cdot 
\hat{u}$, having the form
\begin{eqnarray}
T_{\hat{u}} (\theta, \phi) &=& -\partial E(\vec{M}(\theta,
\phi))/\partial  \theta \;,
\end{eqnarray}
was calculated for $\phi = 0$
(assuming weak anisotropy within the plane perpendicular to the $c$ axis)
and $\theta = \pi/4$, giving immediately the MCA coefficient $K_1$
\cite{SSB+06}.  

In the case of finite temperature, $T > 0$~K, the torque calculations 
for $T_{\hat{u}}(\theta = \pi/4)$  have been performed within the
relativistic disorder local moment (RDLM)
scheme as described by Staunton, et al. \cite{SSB+06,GPS+85}.

In order to investigate the equilibrium magnetic structure and finite
temperature magnetic properties of the compounds under consideration,
Monte Carlo simulations have been performed, which are based on the
extended Heisenberg model with the Hamiltonian accounting for
relativistic effects and represented by the following expression
\begin{eqnarray}
 H &=& 
-\sum_{i,j (i \neq j)}  \hat{e}_i \underline{\underline{J}}_{ij}
 \hat{e}_j 
 +\sum_{i} K_1(\hat{e}_i)_z^2 \;.  
\label{Hspin_2}
\end{eqnarray}
Here $K_1$ is the uniaxial anisotropy constant and
$\underline{\underline{J}}_{ij}$ is the exchange coupling tensor representing
 the isotropic exchange interactions $J_{ij} =
\frac{1}{3}Tr(\underline{\underline{J}}_{ij})$ as well as the components of the
DM interactions  $D^{\lambda}_{ij} =
\frac{1}{2}\epsilon_{\lambda\mu\nu}(J^{\mu\nu}_{ij} - J^{\nu\mu}_{ij})$
\cite{EM09a}.

\section{Results}
\subsection{Structure}
Within the present work we have used structural information taken
from experiment \cite{BC68,HP70,LRI71}.
As it was pointed out in these works, the intercalated 
$M_x$NbS$_2$ systems with $M = $Cr, Mn, and Fe 
exhibit at $x = 1/3$ a $\sqrt{3}\times\sqrt{3}$ superstructure within 
the $M$ layers, leading to a well defined ordered compound
crystallizing in the space group P$6_322$. 
This implies an occupation of the $(12i)$ Wyckoff positions (with
$x=1/3, y=0, z=3/8$) by S atoms, and occupation of the $2a$ and $4f$
(with $z=0$) positions by Nb atoms. For the $M$ atoms, this symmetry
group allows three possible positions: $2b$, $2c$ and $2d$. 
Experimental results, however, suggest the occupation
of the $2c$ position (1/3,2/3,1/4) to be most probable. 
In order to investigate the impact of the optimization of structure
parameters via DFT calculations, a corresponding optimization has been 
performed using the Vienna ab initio simulation package (VASP)
\cite{KF96,KF96a}. The optimized structure parameters for 
the compounds with the  $M$ atoms occupying the $2c$ Wyckoff positions
are presented in Table \ref{TAB_STRUC} together with the 
experimental data.

\begin{table}

\begin{tabular}{ |c|c|c|c|c|  }
 \hline
 \multicolumn{1}{|c|}{}   &\multicolumn{2}{|c|}{Expt} & \multicolumn{2}{|c|}{Theory}  \\
 \hline
      & a  & c  & a & c  \\
 \hline
   Cr$_{1/3}$NbS$_2$  & 5.73   &  12.11   &  5.782  &  12.141  \\
   Mn$_{1/3}$NbS$_2$  & 5.779    & 12.599   &  5.814  &  12.378  \\
   Fe$_{1/3}$NbS$_2$  & 5.761    & 12.178   &  5.794  &  12.084  \\
\hline
\end{tabular}
\caption{\label{TAB_STRUC} The structure parameters (in \AA) for Cr-
  \cite{HP70}, Mn- \cite{LRI71}, and Fe-  \cite{LRI71} intercalated  NbS$_2$ for
$M$ occupying the (2c) Wyckoff position.}     
\end{table}

\subsection{Electronic structure}

\begin{figure}[h]
\includegraphics[width=0.4\textwidth,angle=0,clip]{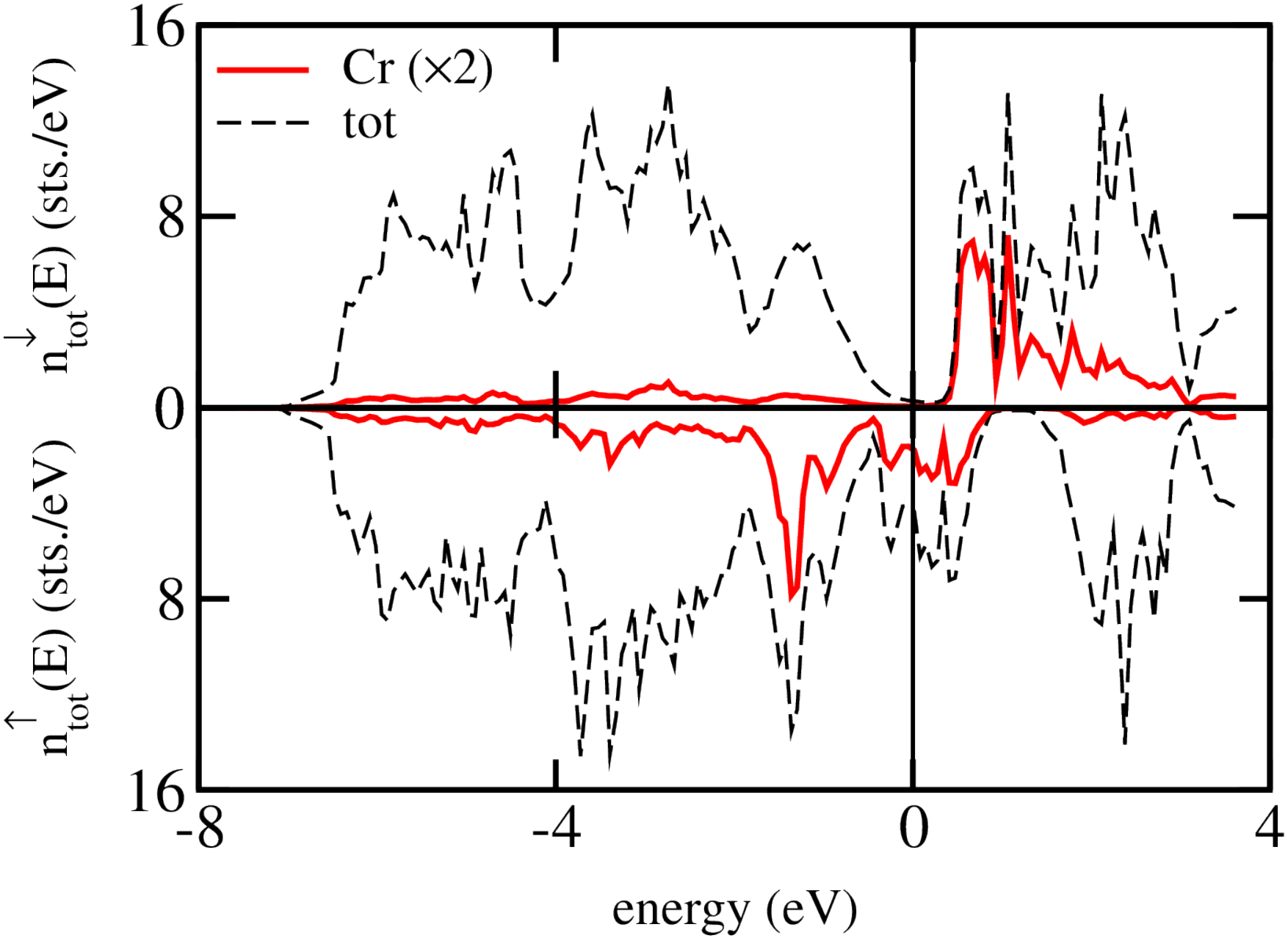}\;
(a) \\
\includegraphics[width=0.4\textwidth,angle=0,clip]{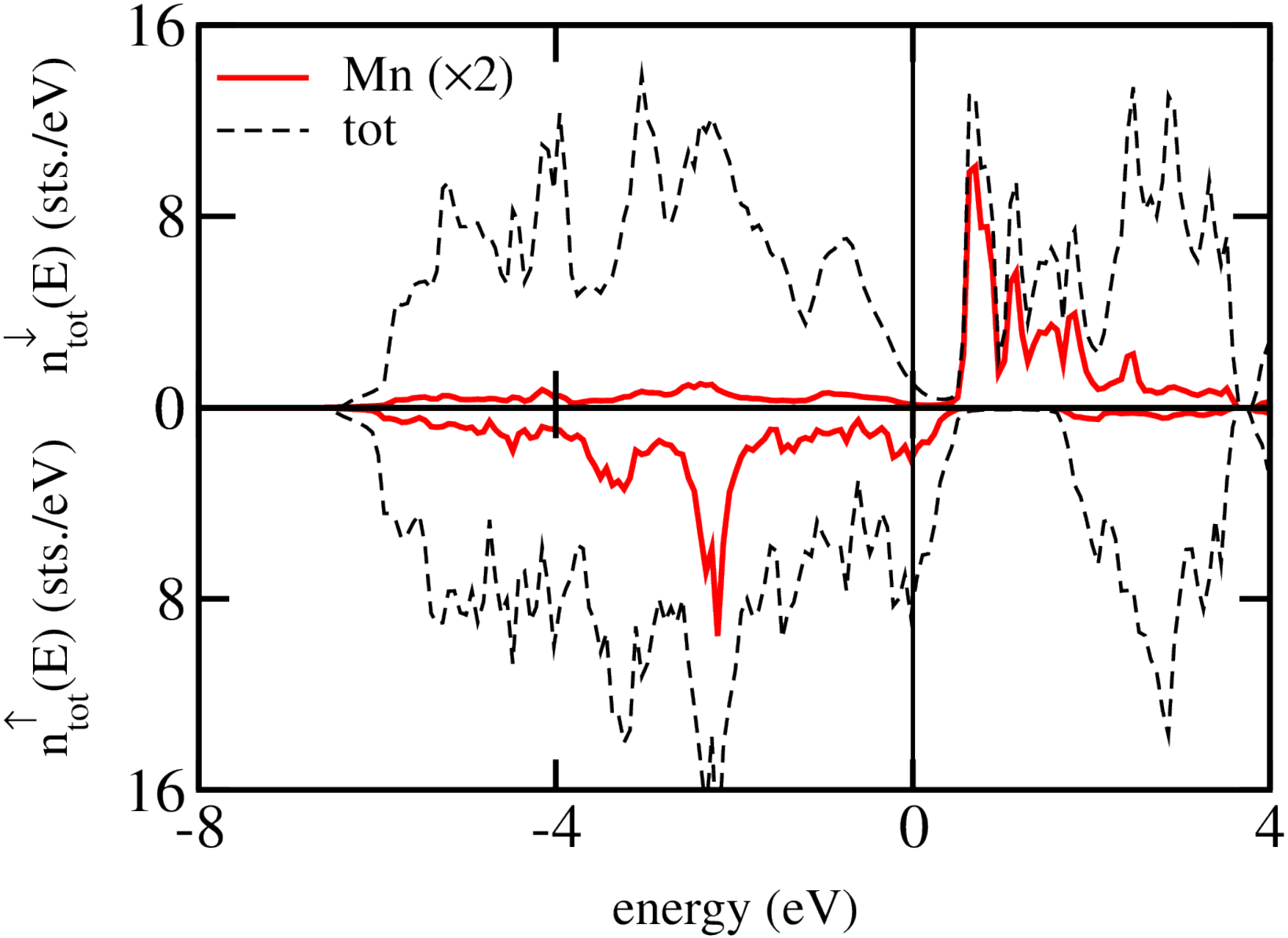}\;
(b) \\
\includegraphics[width=0.4\textwidth,angle=0,clip]{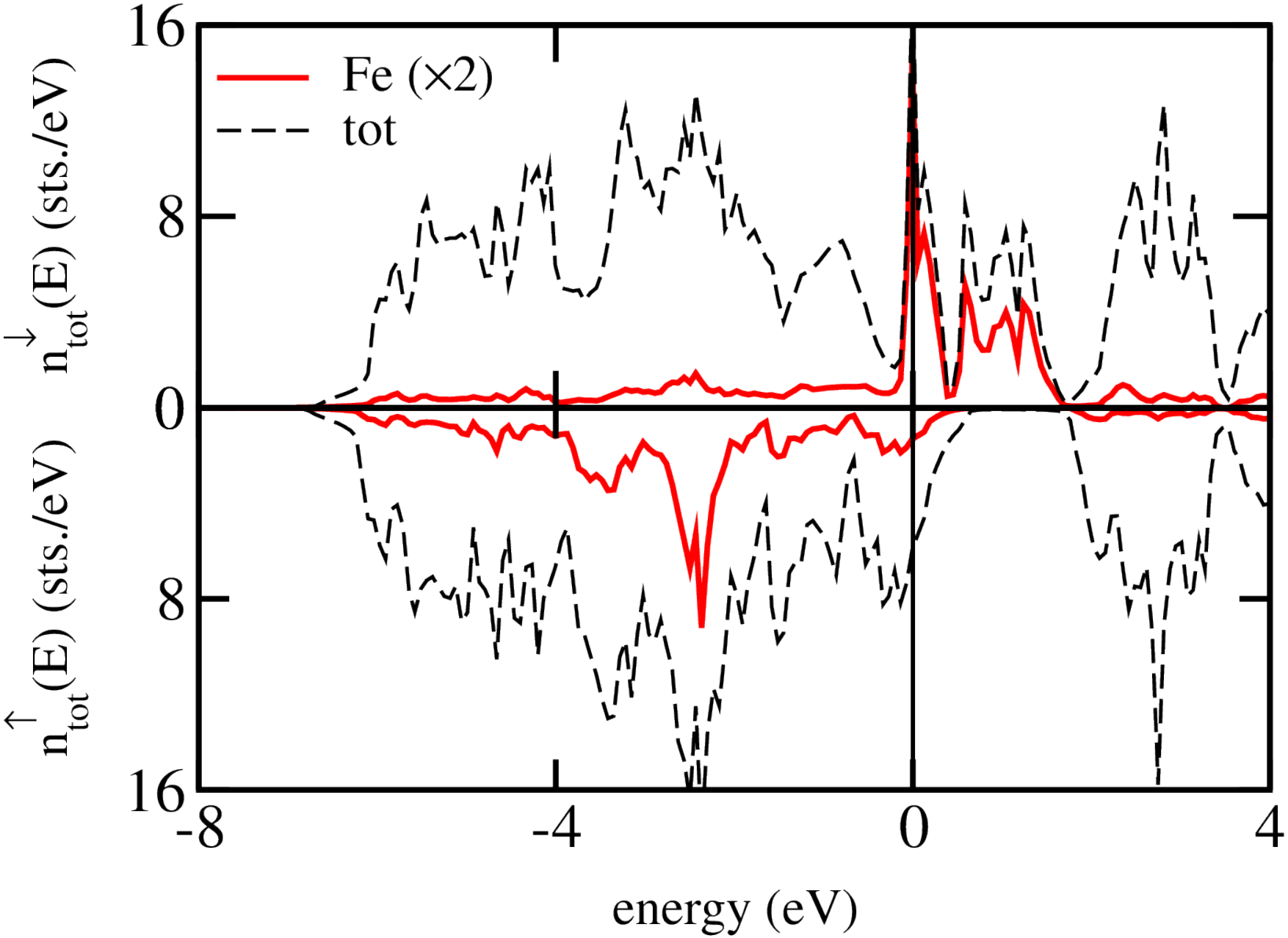}\;
(c) 
\caption{\label{fig:CMP_TOTDOS} The total DOS for the FM-ordered
  $M_{1/3}$NbS$_2$ ($M$ = Cr, Mn, Fe) together with the partial DOS for
  the $M$ atom (scaled by the factor of 2). }  
\end{figure}

\begin{figure}[h]
\includegraphics[width=0.25\textwidth,angle=270,clip]{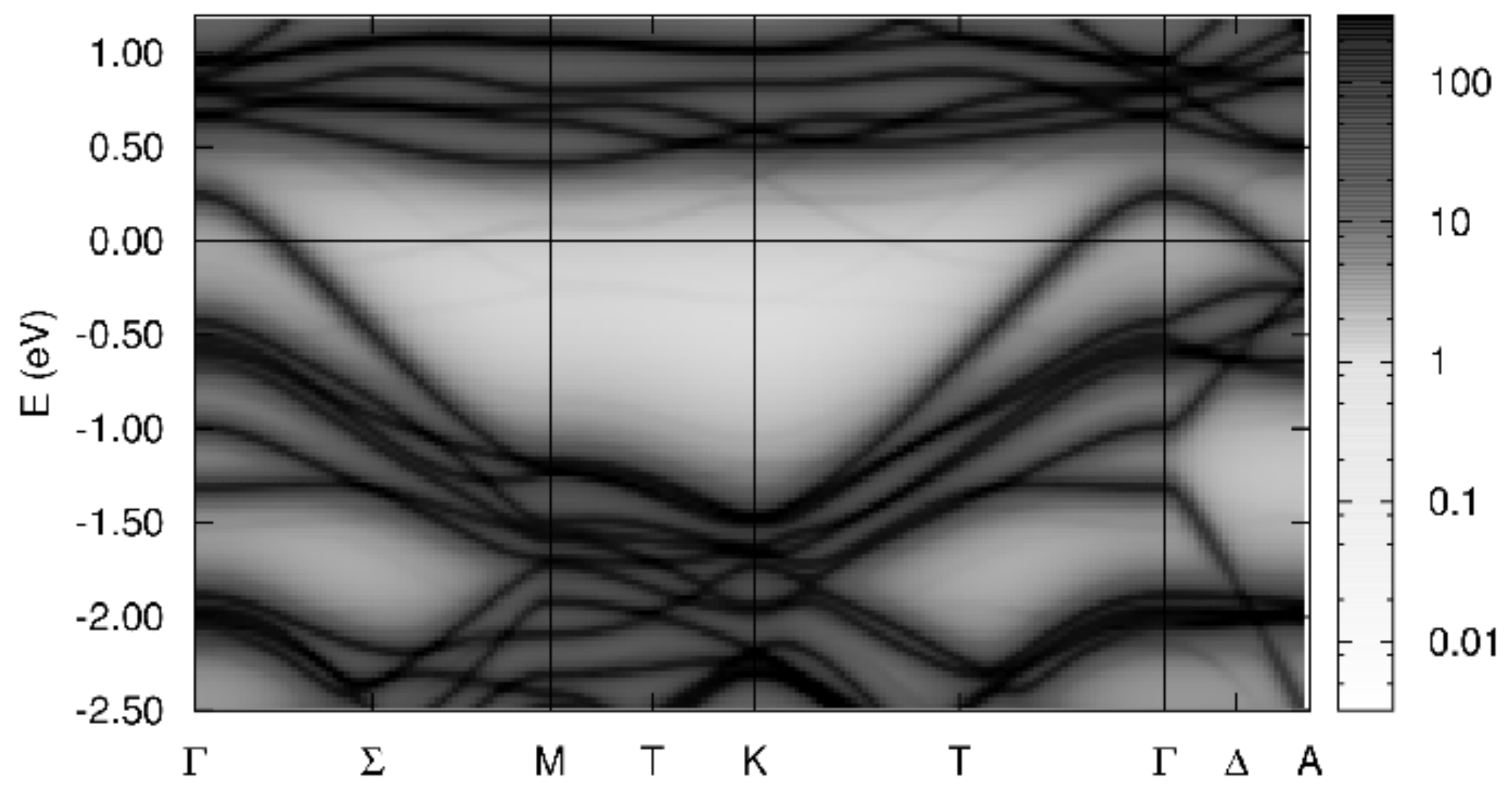}\; (a)\\
\includegraphics[width=0.25\textwidth,angle=270,clip]{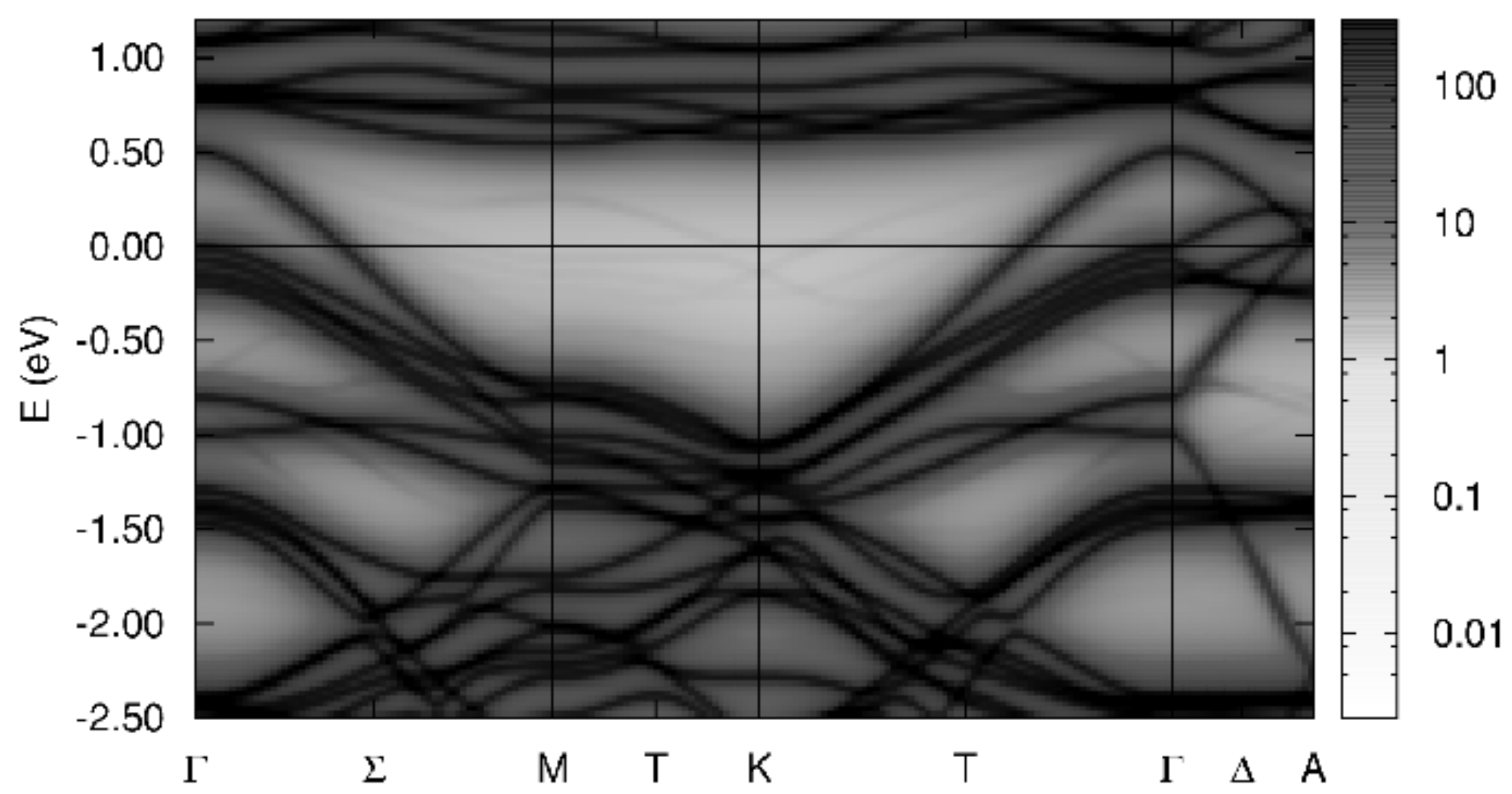}\; (b)\\
\includegraphics[width=0.25\textwidth,angle=270,clip]{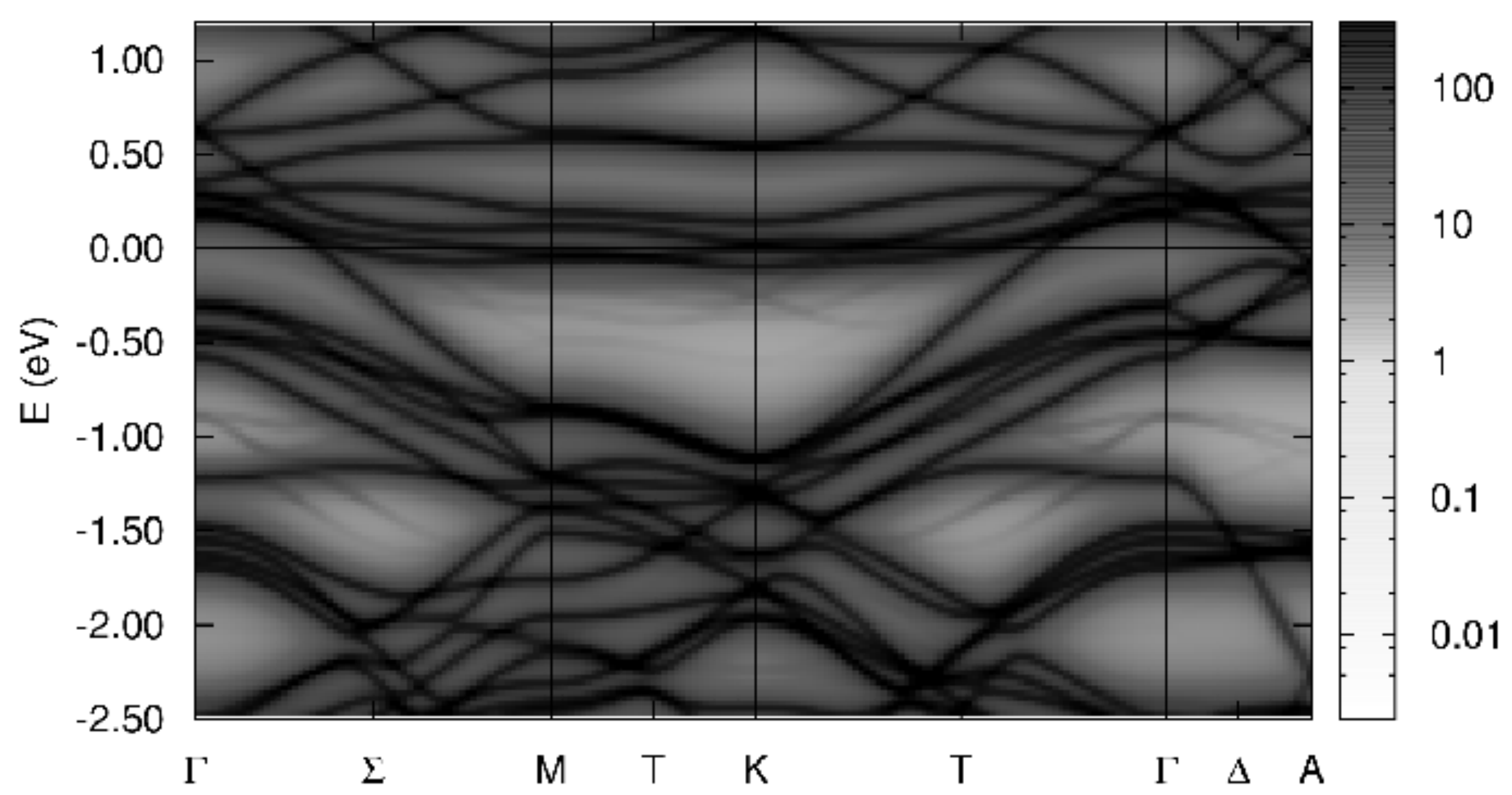}\; (c)
\caption{\label{fig:Fe_BSF_dw} Spin-down  Bloch spectral functions for the
  FM-ordered  Cr$_{1/3}$NbS$_2$  (a) Mn$_{1/3}$NbS$_2$   (b) and
  Fe$_{1/3}$NbS$_2$    (c), calculated using an imaginary part for the
  energy of $1$~meV.  }    
\end{figure}

The electronic structure calculations have been performed for the FM
state for all systems under consideration. The total spin projected
densities of states (DOS), $n(E)$, are represented in
Fig.\ \ref{fig:CMP_TOTDOS}. As one can see, the DOS of the majority-spin
states at the Fermi energy $E_F$ is rather large for all  
compounds. In the case of minority-spin states of  Cr$_{1/3}$NbS$_2$  and  Mn$_{1/3}$NbS$_2$
one can see a 'pseudogap' between the occupied and unoccupied states
with a finite DOS at the Fermi level originating from the
$d$-states of Nb crossing the Fermi level (see the Bloch spectral function 
(BSF) in Fig. \ref{fig:Fe_BSF_dw}). In the case of 
Fe$_{1/3}$NbS$_2$ the Fermi energy is located at the DOS maximum
corresponding to the Fe minority-spin $d$-states and is finite for both spin
directions $n(E_F)$.

Table \ref{TAB_MOM} represents the spin and orbital magnetic moments
of the $M$ atoms in comparison with experimental results derived from
measurements of the saturation magnetization \cite{HP70}. The calculated spin magnetic
moments are obviously in good agreement with the experimental data.

\begin{table}

\begin{tabular}{ |c||c|c|c|c|c|c|  }
 \hline
               & $m_{spin}$ & $m_{orb}$ & $m^o_{spin}$ & $m^o_{orb}$ &
               $m^{expt}_{spin}$ & $E_{MCA}$  \\
 \hline
  Cr$_{1/3}$NbS$_2$  & 2.916   & -0.022 & 2.914   & -0.023   &  2.9 & 0.21  \\
  Mn$_{1/3}$NbS$_2$  & 3.791   &  0.014 & 3.758   & 0.013    &  3.8 & 0.48  \\
  Fe$_{1/3}$NbS$_2$  & 2.905   &  0.215 & 2.924   & 0.210    &   -  & -1.56 \\
\hline
\end{tabular}
\caption{\label{TAB_MOM} Spin and orbital magnetic moments of $M$
  atoms in $\mu_B$/atom: present results vs experiment (derived from
  saturation measurements) \cite{HP70,MKK+83}. $m^o_{spin}$ and $m^o_{orb}$
  represent results obtained using the structure parameters for
  the optimized structure (see Table \ref{TAB_STRUC}). The calculated MCA energy
  $E_{MCA}$ is represented in meV/(f.u.). A negative value implies the
  easy-axis concerning the MCA to be along the $c$ axis.}       
\end{table}

\begin{figure}[h]
\includegraphics[width=0.25\textwidth,angle=270,clip]{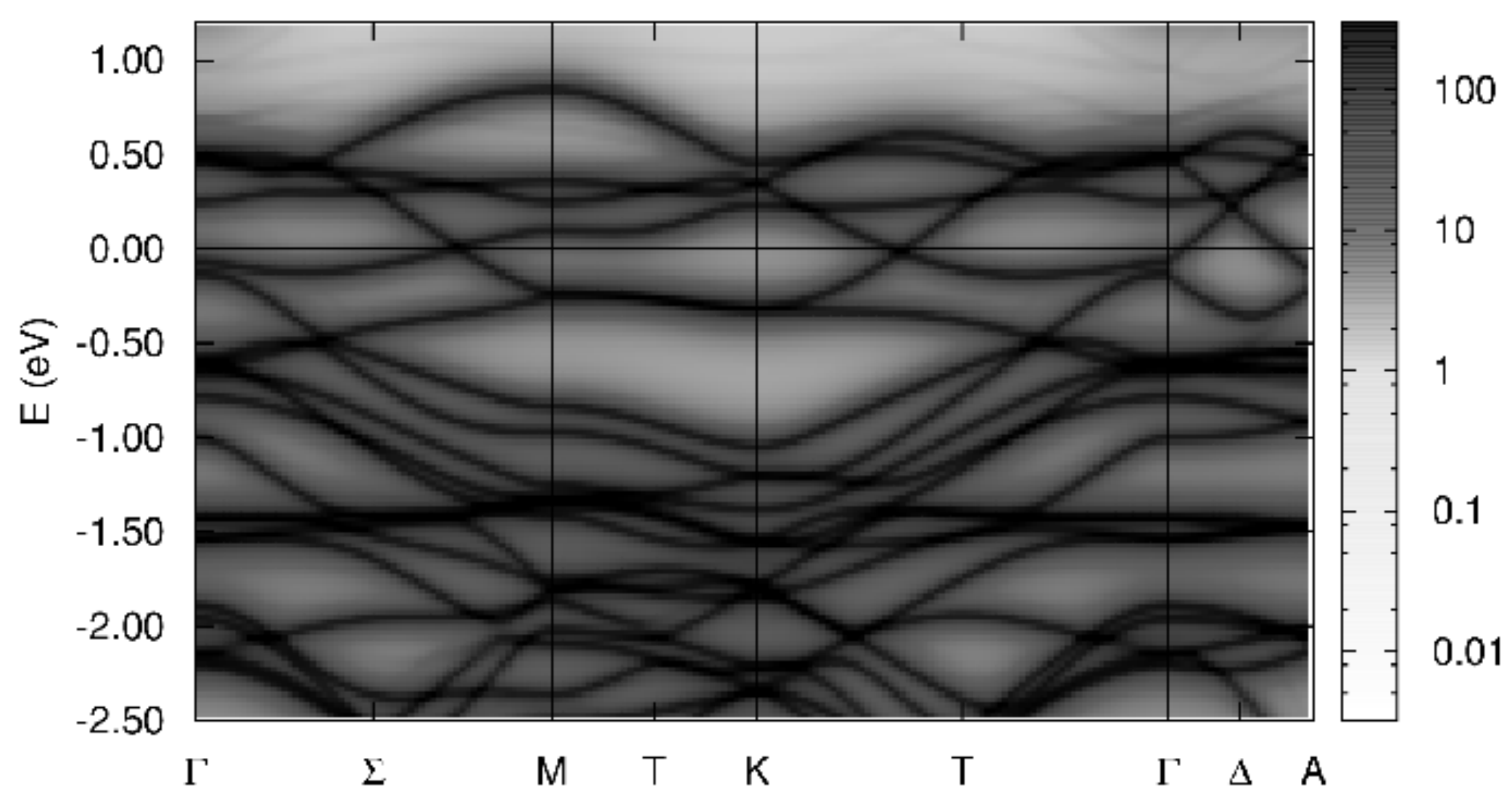}\; (a)\\
\includegraphics[width=0.25\textwidth,angle=270,clip]{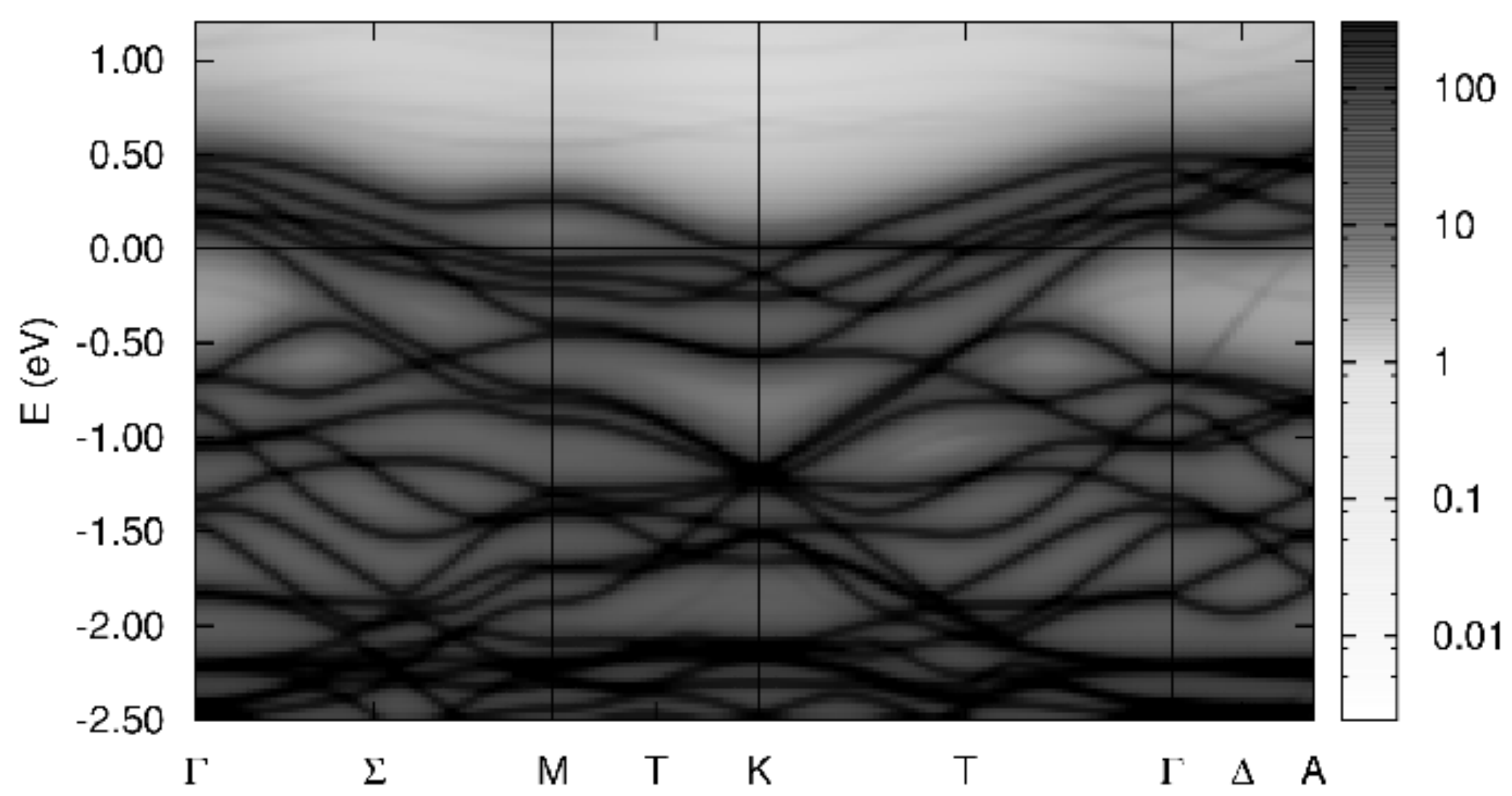}\; (b)\\
\includegraphics[width=0.25\textwidth,angle=270,clip]{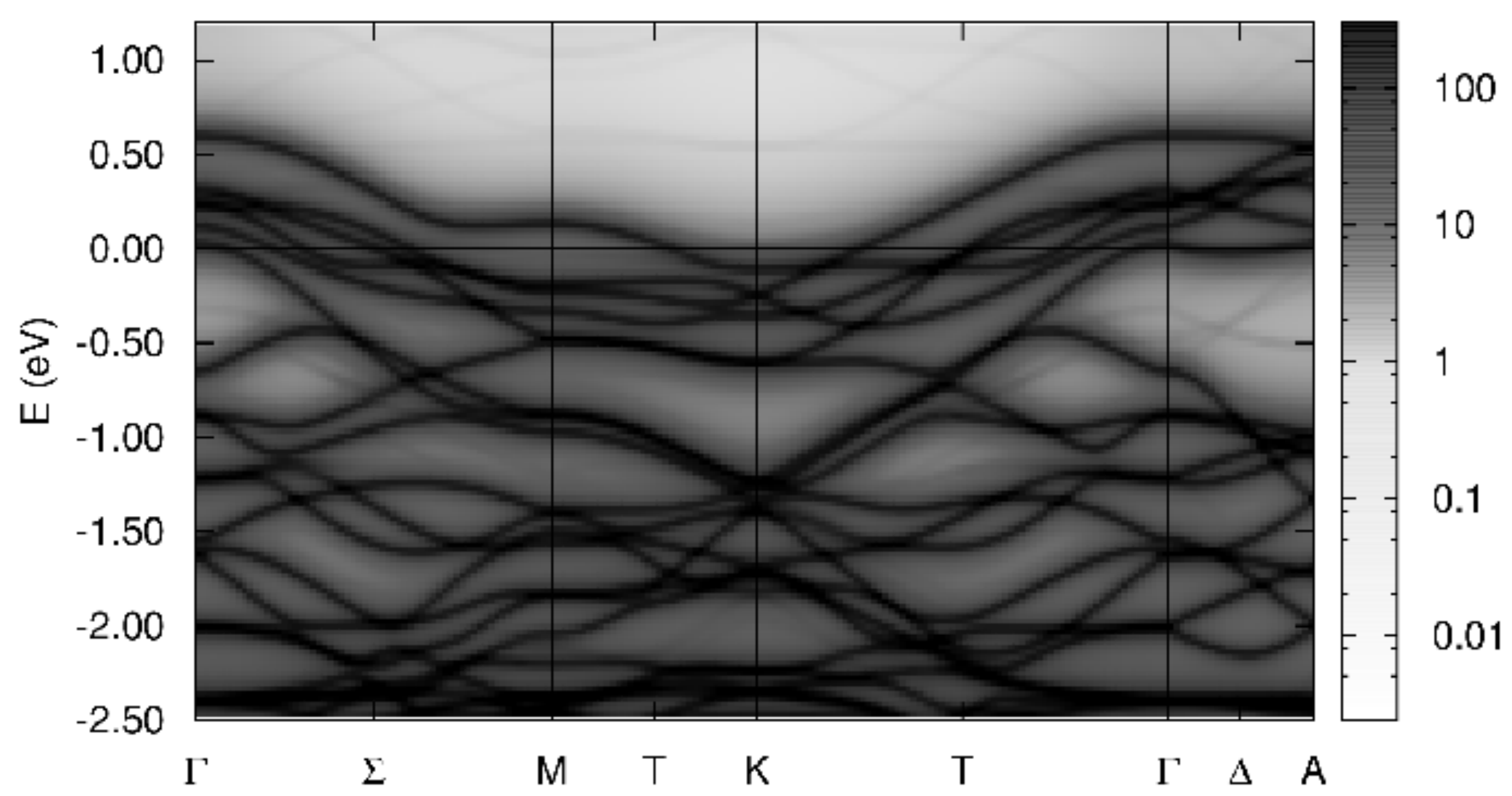}\; (c)
\caption{\label{fig:Fe_BSF_up} Spin-up  Bloch spectral functions for the
  FM-ordered  Cr$_{1/3}$NbS$_2$  (a) Mn$_{1/3}$NbS$_2$   (b) and
  Fe$_{1/3}$NbS$_2$    (c), calculated using an imaginary part for the
  energy of $1$~meV. }   
\end{figure}

One can also see that the Cr and Mn orbital magnetic moments in the 
corresponding intercalated systems are rather small. In contrast, Fe
in Fe$_{1/3}$NbS$_2$ has an orbital magnetic moment that is larger by an order of
magnitude. This can be attributed to the narrow-band of the Fe minority-spin
$d$-states at the Fermi energy (see Figs. \ref{fig:CMP_TOTDOS} and
\ref{fig:Fe_BSF_dw}).    
In Table \ref{TAB_MOM} we represent also the
spin and orbital magnetic moments obtained for the systems using the
optimized structure parameters shown in Table \ref{TAB_STRUC}. As one
can see, they are both close to the experimental values as well as to
the results obtained for the systems with experimental structure
parameters. 
In the last column of the Table \ref{TAB_MOM} results for the
MCA are shown calculated for the FM ordered $M_{1/3}$NbS$_2$ systems for $T =
0$~K. These results imply an uniaxial easy plane  
anisotropy in the case of Cr$_{1/3}$NbS$_2$ and Mn$_{1/3}$NbS$_2$
and an easy axis along the $c$ axis in the case of
Fe$_{1/3}$NbS$_2$, fully in line with experiment. 
Note also that the magnitude of the MCA in the former cases is much smaller
than in the latter one. This can be easily understood by analyzing in
detail the electronic structure represented in 
Figs. \ref{fig:Fe_BSF_up} and \ref{fig:Fe_BSF_dw}. 
%
%
%
For the case of Fe$_{1/3}$NbS$_2$ the band with 
minority spin $d$-character is rather narrow and is located at the
Fermi level. This results in a significant contribution to the energy
$\sim w^{-1}$ due to SOC splitting of the states having the same spin
direction, with the bandwidth $w$, leading to 
a preferable magnetization direction along the quantization axis
parallel to the hexagonal $c$ axis (see, e.g. the
discussions by Wu et al. \cite{WWF93}).
In the case of Cr$_{1/3}$NbS$_2$ and Mn$_{1/3}$NbS$_2$ the
Fermi level is positioned between the spin-up and spin-down $d$-bands of
the $3d$ metal Cr and Mn, respectively. This results in a significant SOC-induced energy gain
associated with the states having opposite spin directions, which
competes with the spin-diagonal part of the energy. 
Using the average exchange splitting $\Delta_{ex}$ for the Cr (Mn)
$d$-states, the SOC-induced energy gain for the in-plane orientation of
the magnetic moment is $\sim \Delta_{ex}^{-1}$, which is comparable with the
energy gain for $\vec M$ oriented along $c$, and $\sim w_{up}^{-1}$.
As given by Table \ref{TAB_MOM} these lead for these two
compounds to an easy plane MCA. The competition of two contributions to
the MCA energy results in a smaller magnitude of MCA of Cr$_{1/3}$NbS$_2$
and Mn$_{1/3}$NbS$_2$ when compared to the Fe$_{1/3}$NbS$_2$.

\begin{figure}[h]
\includegraphics[width=0.45\textwidth,angle=0,clip]{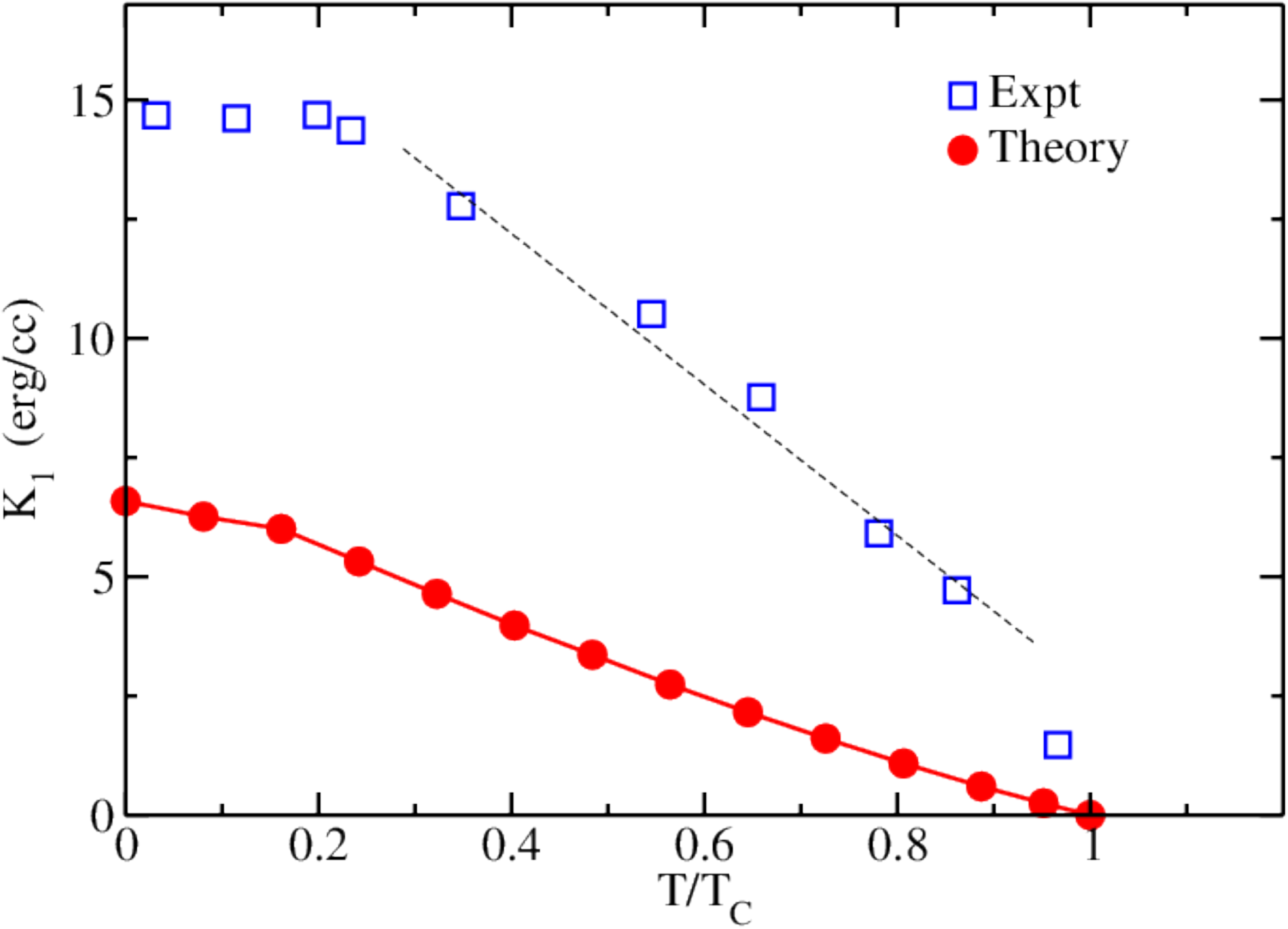}
\includegraphics[width=0.45\textwidth,angle=0,clip]{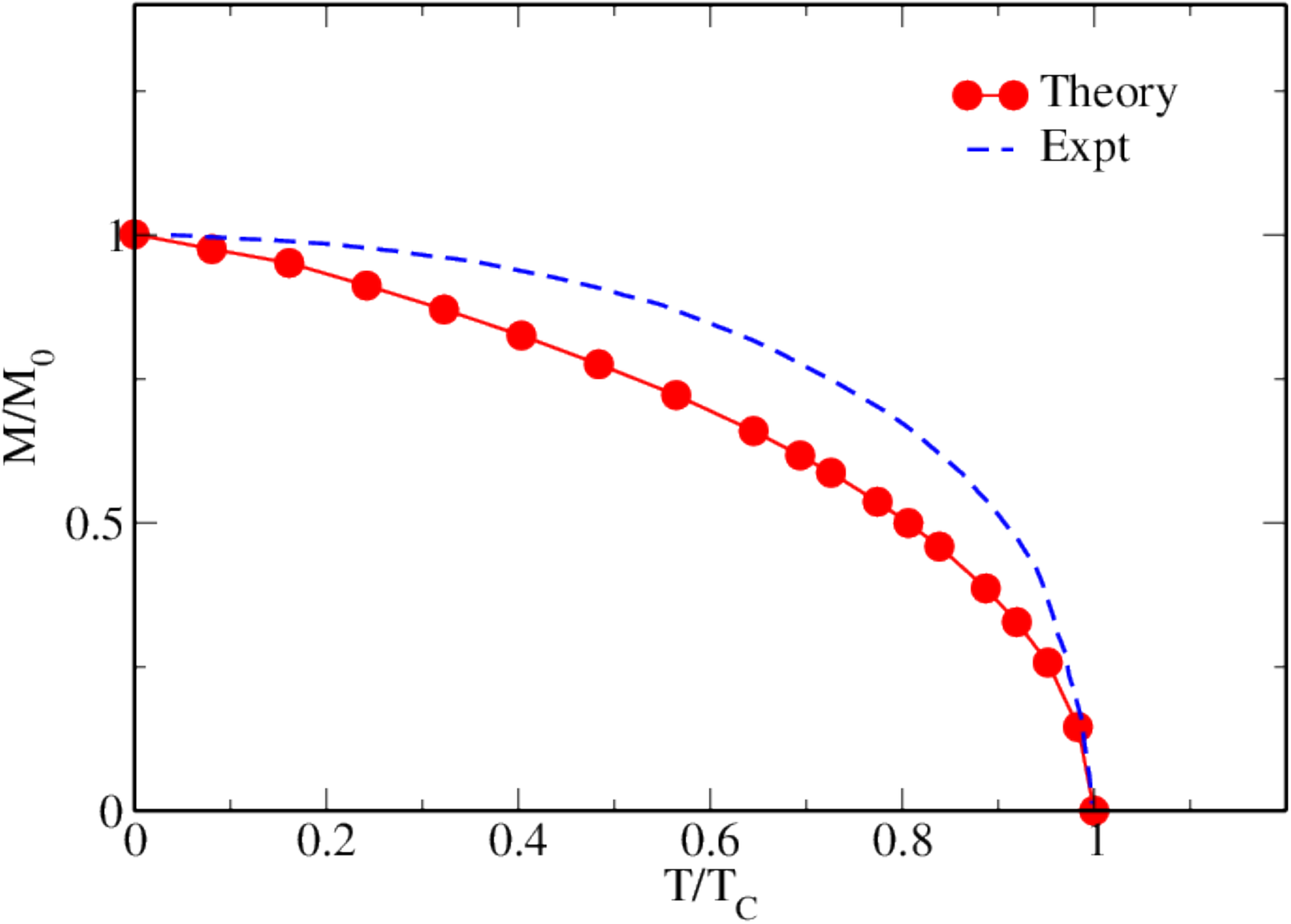}
\caption{\label{fig:MCA_CrNbS2} (a) Magneto-crystalline anisotropy in
  Cr$_{1/3}$NbS$_2$ as a function of temperature obtained via RDLM
  selfconsistent calculations in comparison with experimental
  results \cite{MKK+83}. (b) Temperature-dependent variation of the magnetization
  calculated within the mean field approximation (RDLM scheme) in comparison with the
  experimental results obtained using the external magnetic field  \cite{MKK+83}.  }  
\end{figure}

The temperature dependence of the MCA  for
Cr$_{1/3}$NbS$_2$ has been investigated within  RDLM
calculations. Figure \ref{fig:MCA_CrNbS2}(a) represents the 
coefficient of the uniaxial magnetic anisotropy, $K_1$, as a
function of temperature, in comparison with experimental
results. One can see a nearly linear dependence of the MCA energy on the
temperature, in full agreement with experiment 
 \cite{MKK+83}. Although the calculations underestimate
the experiment by about a factor of two, one can nevertheless call the 
agreement between theory and experiment very satisfying. Figure \ref{fig:MCA_CrNbS2}(b)
displays also the temperature-dependent variation of the magnetization
obtained via RDLM calculations, in comparison with experiment. As one notices, 
the theoretical $M(T)$ curve drops faster with increasing $T$ as the 
experimental one as the RDLM does not include long-range magnetic correlations.
%

\begin{figure}[h]
\includegraphics[width=0.35\textwidth,angle=0,clip]{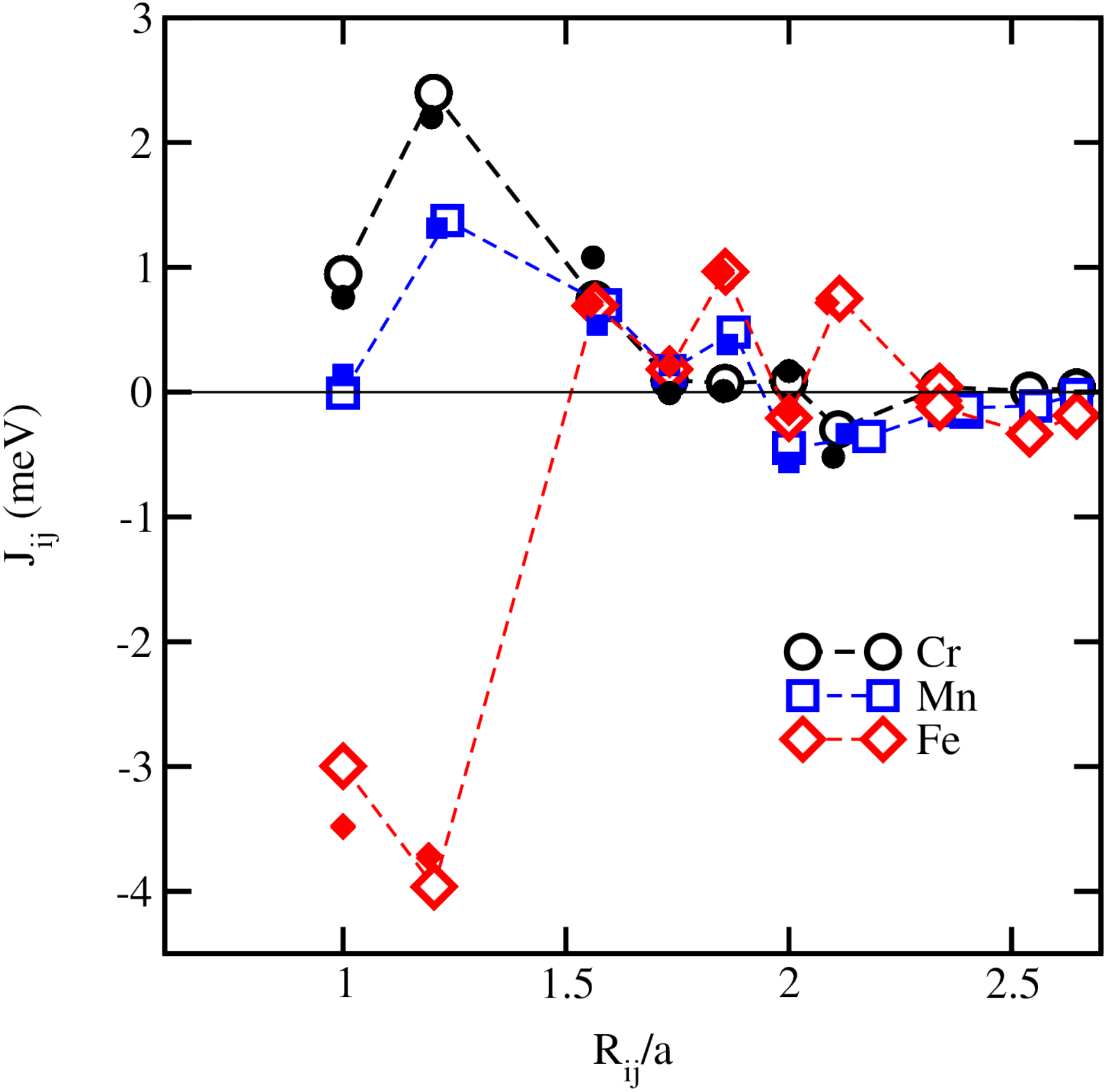}\; (a)
\includegraphics[width=0.35\textwidth,angle=0,clip]{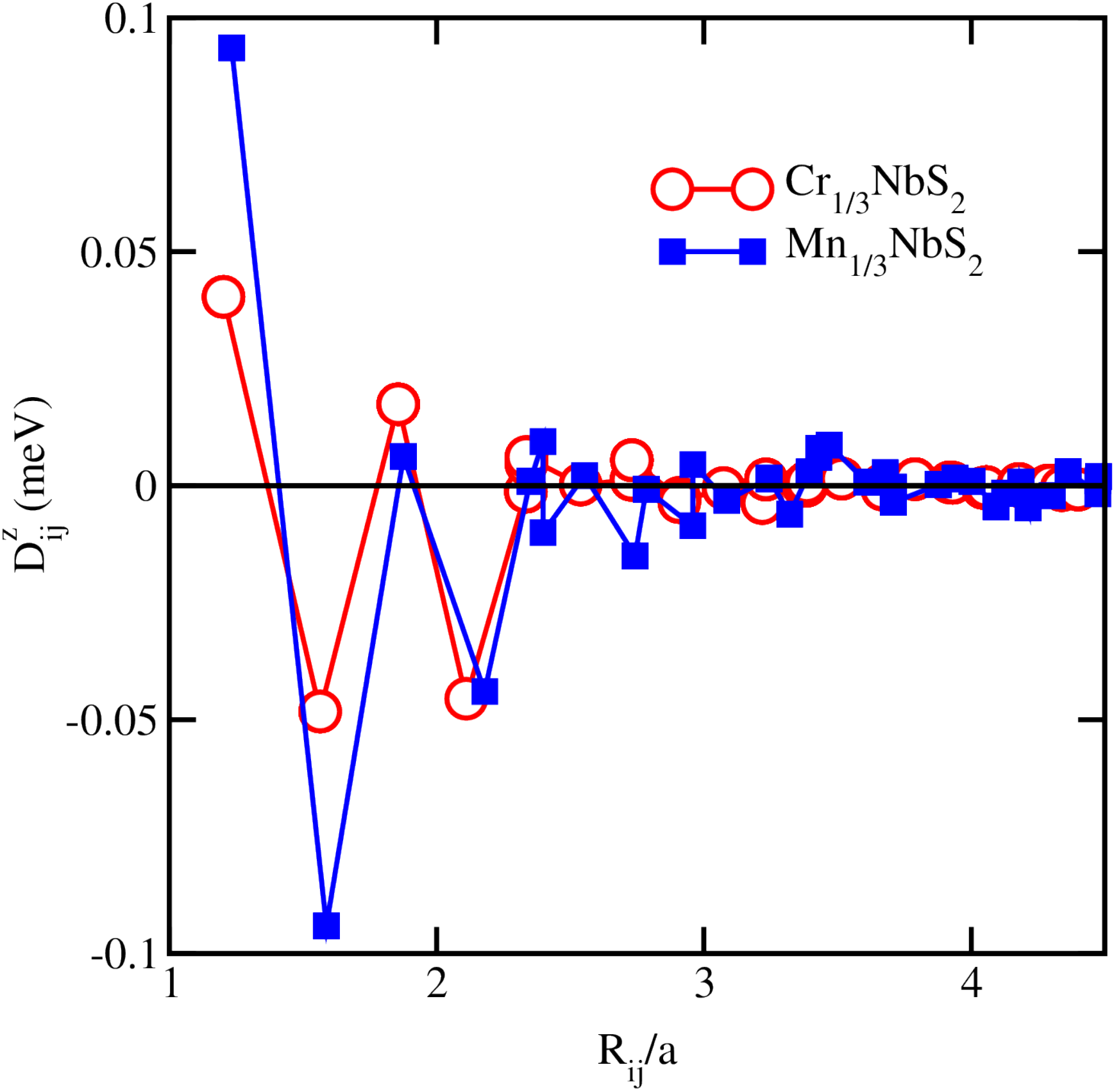}\;(b)
\caption{\label{fig:CMP_JXC} (a) Isotropic exchange coupling parameters for
  $M_{1/3}$NbS$_2$ ($M$ = Cr, Mn, Fe). Open symbols represent results
  obtained for the experimental structure parameters, full symbols
  correspond to the systems with the DFT-optimized structure parameters.
(b) The $D^z$ component of the DM interactions of Cr and Mn atoms with
the neighbours arranged within the layers corresponding to the
bottom half-space of the system. }   
\end{figure}

Figure \ref{fig:CMP_JXC} represents the  isotropic exchange
coupling parameters $J_{ij}$ calculated for all three systems. The
strongest magnitude are found for the 
$M$-$M$ interactions for the first two neighbour shells. They are
predominantly positive for the case of Cr$_{1/3}$NbS$_2$  and
Mn$_{1/3}$NbS$_2$ compounds, leading to FM order 
in these systems when the Dzyaloshinskii-Moriya (DM) interactions are not
taken into account. In the case of Fe$_{1/3}$NbS$_2$, however, the Fe-Fe
interactions at short distances are negative. Due to the hexagonal crystal
structure this should lead to a frustrated magnetic structure. 
The exchange coupling parameters calculated for the systems with
optimized structure parameters are also presented in
Fig. \ref{fig:CMP_JXC}  (full symbols), demonstrating a rather
small difference from the results obtained for experimental structure
parameters (open symbols).
%

\begin{figure}[h]
 \includegraphics[width=0.15\textwidth,angle=0,clip]{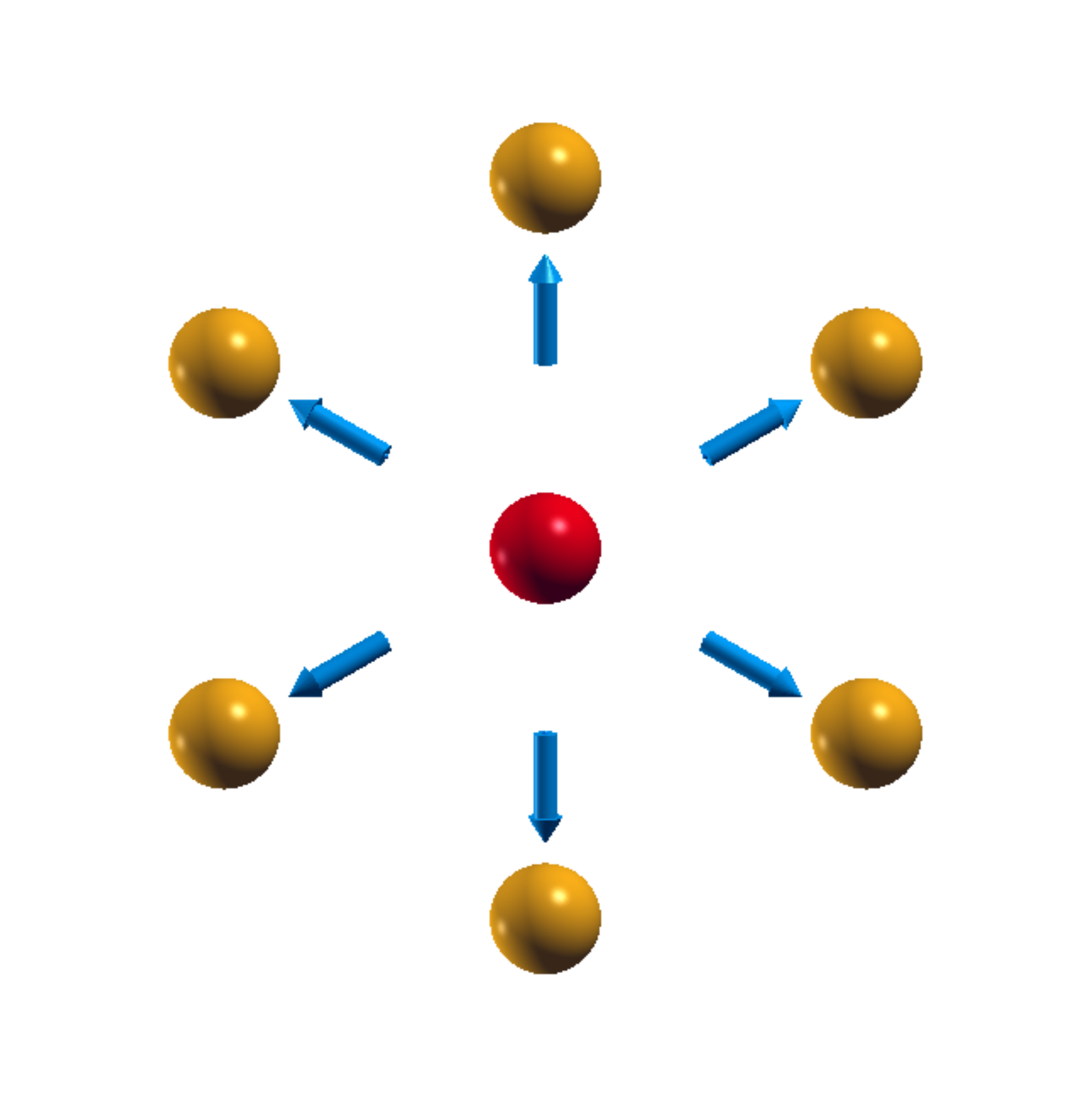} 
\\
Cr:\;\;\;$|\vec{D}_{01}| = 0.17 meV$ \;\;\;\;\;$|\vec{D}_{04}| = 0.01 meV$ \\
Mn:\;\;\;$|\vec{D}_{01}| = 0.09 meV$ \;\;\;\;\;$|\vec{D}_{04}| = 0.01 meV$ \\
Fe:\;\;\;$|\vec{D}_{01}| = 1.02 meV$ \;\;\;\;\;$|\vec{D}_{04}| = 0.03
meV$ \\
\; \; \;(a) \\
\includegraphics[width=0.15\textwidth,angle=0,clip]{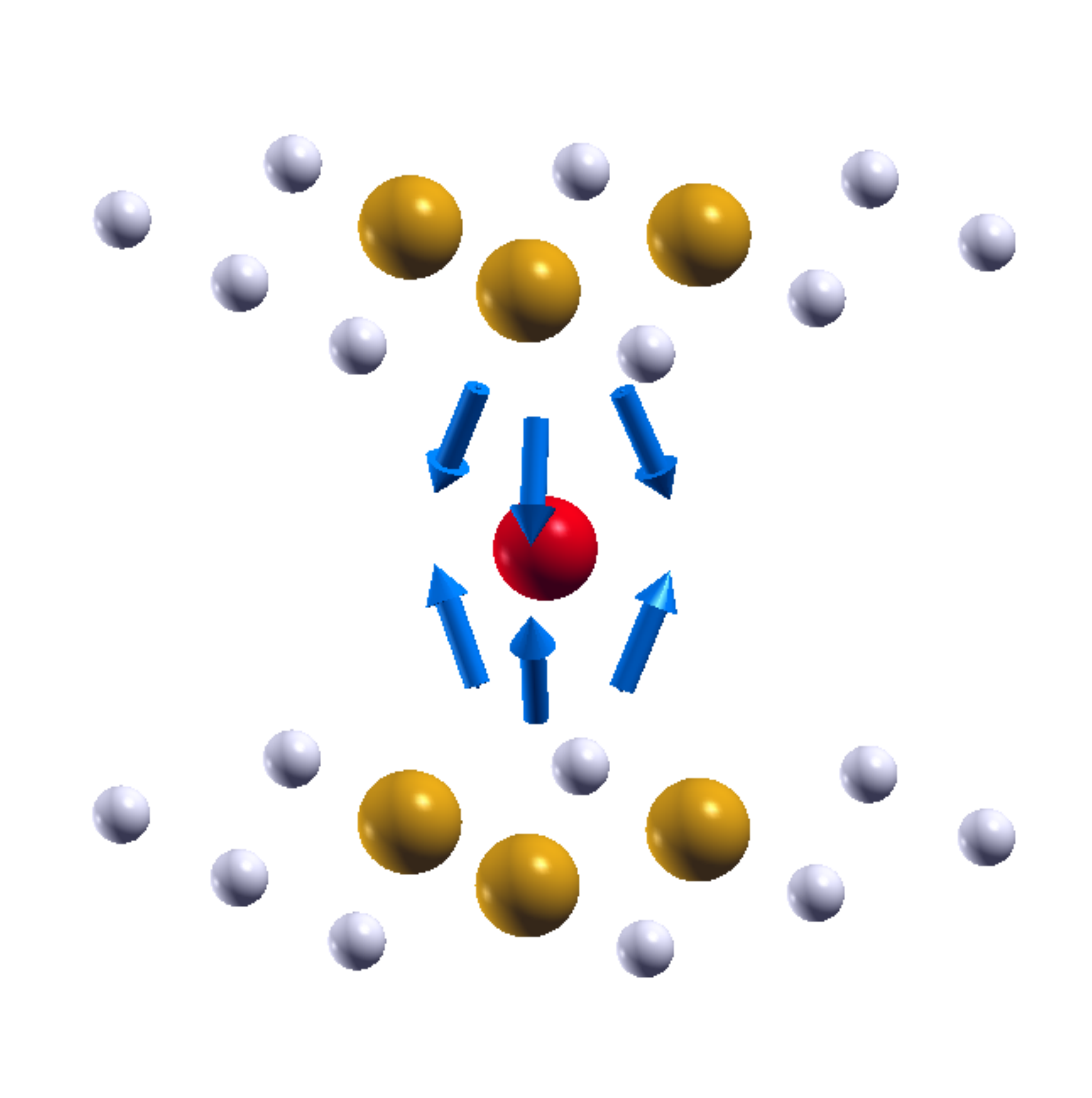}
\includegraphics[width=0.15\textwidth,angle=0,clip]{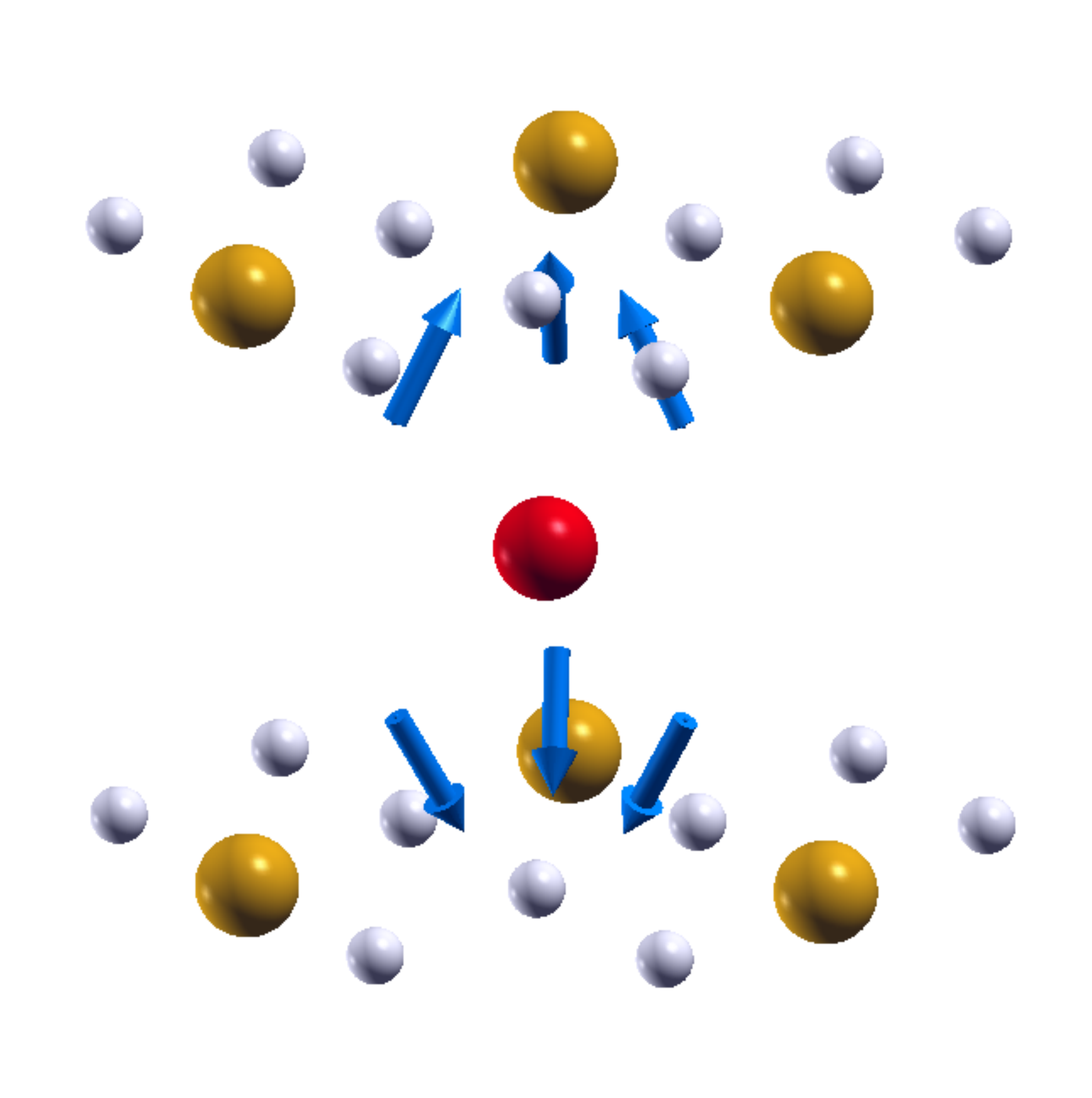}
\includegraphics[width=0.15\textwidth,angle=0,clip]{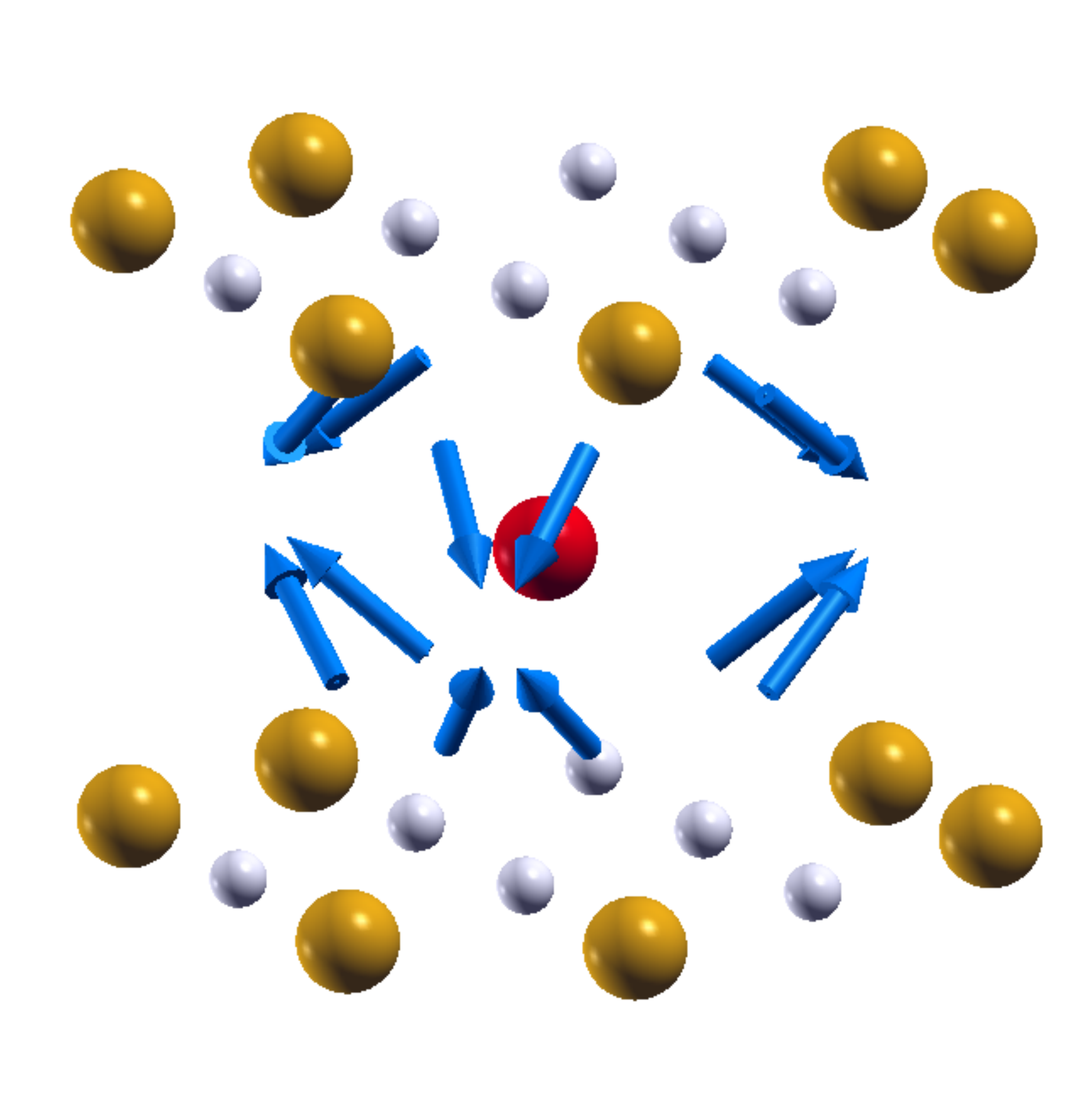}
\\
$|\vec{D}_{02}| = 0.04 meV$ \;\;\;\;\;$|\vec{D}_{03}| = 0.06 meV$
\;\;\;\;\;$|\vec{D}_{05}| = 0.021 meV$ \\
$|D^z_{02}| = 0.04 meV$ \;\;\;\;\;$|D^z_{03}| = 0.05 meV$
\;\;\;\;\;$|D^z_{05}| = 0.017 meV$ \\\; \; \;(b)
\\
\includegraphics[width=0.15\textwidth,angle=0,clip]{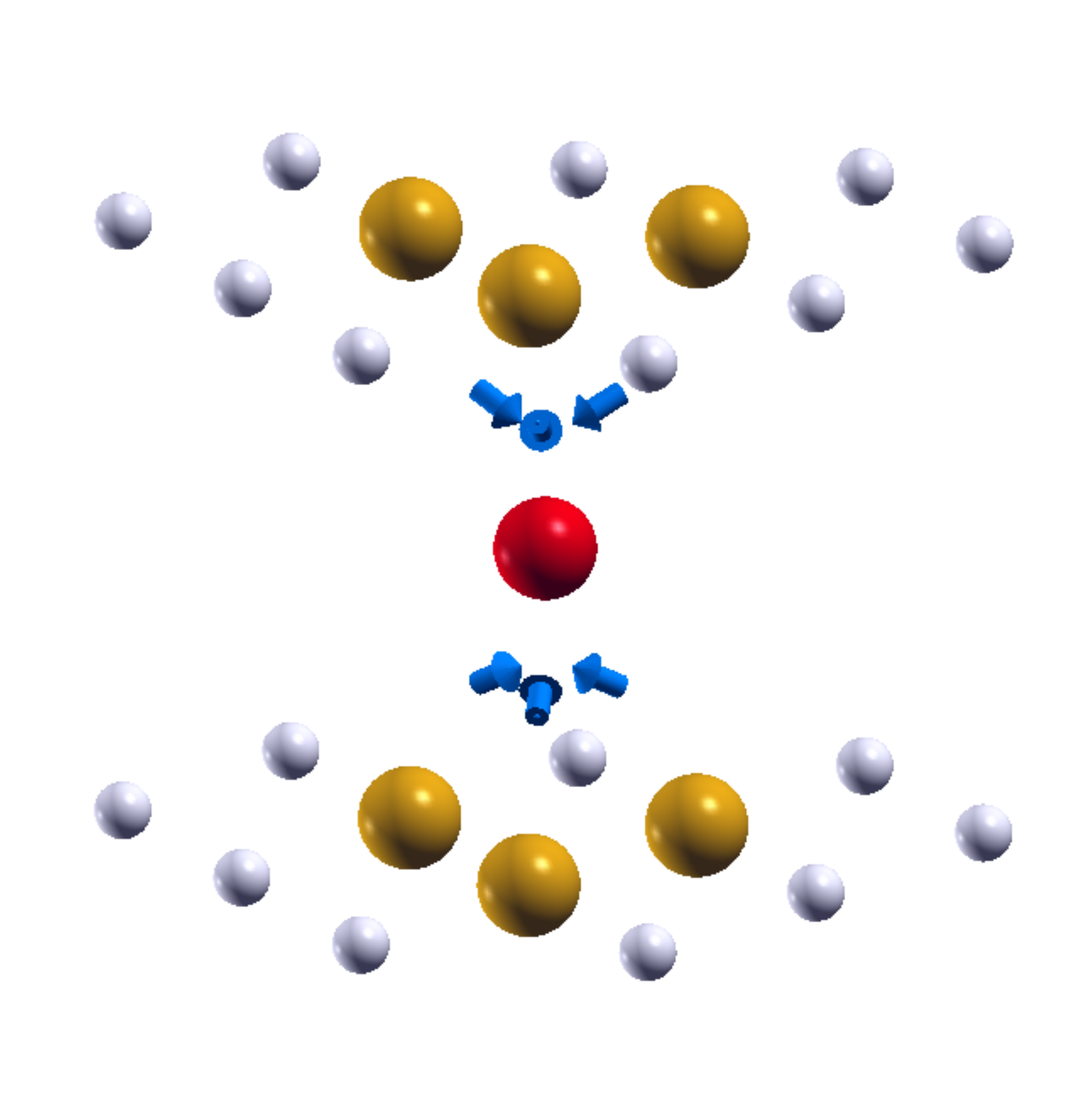}
\includegraphics[width=0.15\textwidth,angle=0,clip]{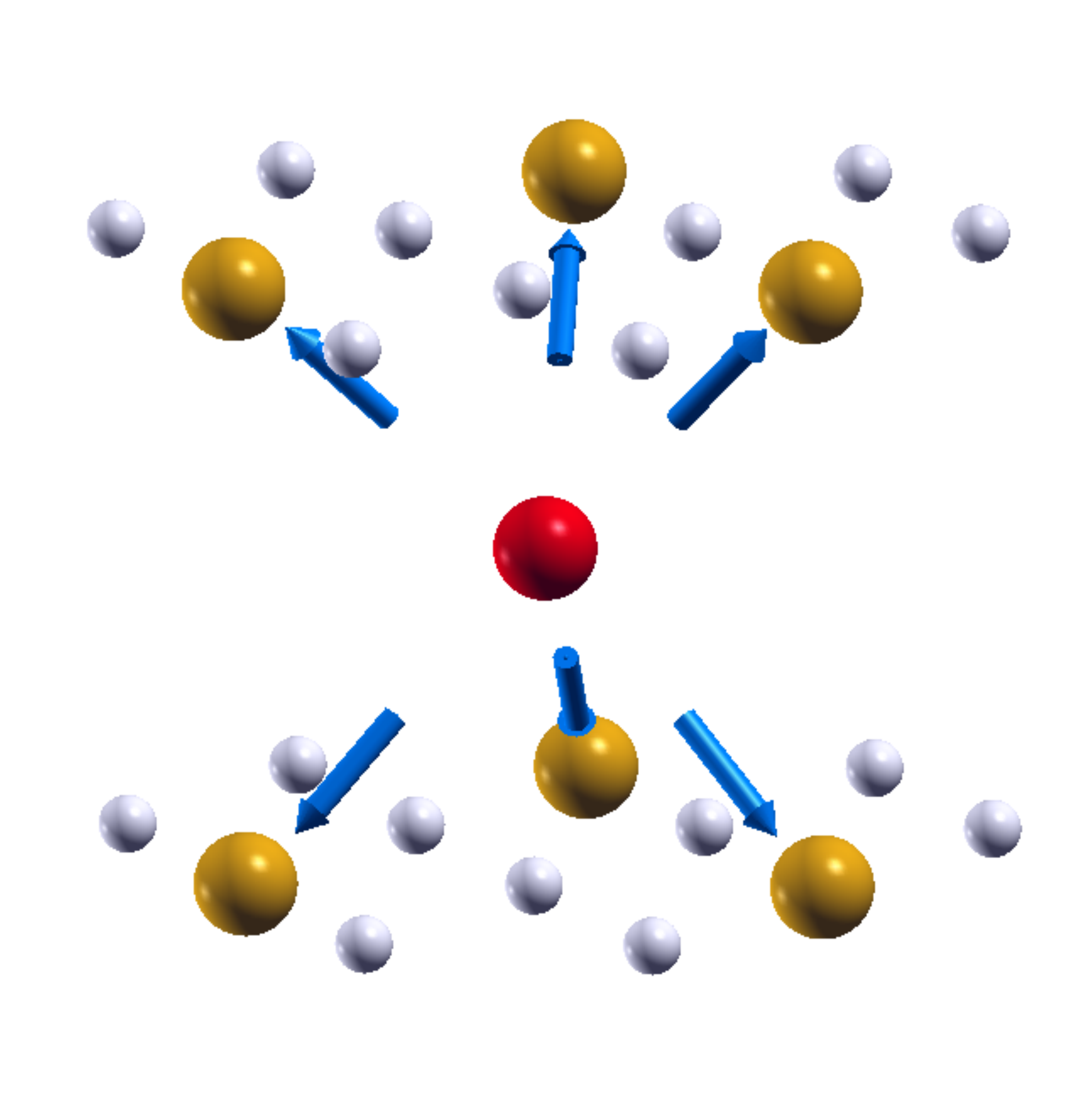}
\includegraphics[width=0.15\textwidth,angle=0,clip]{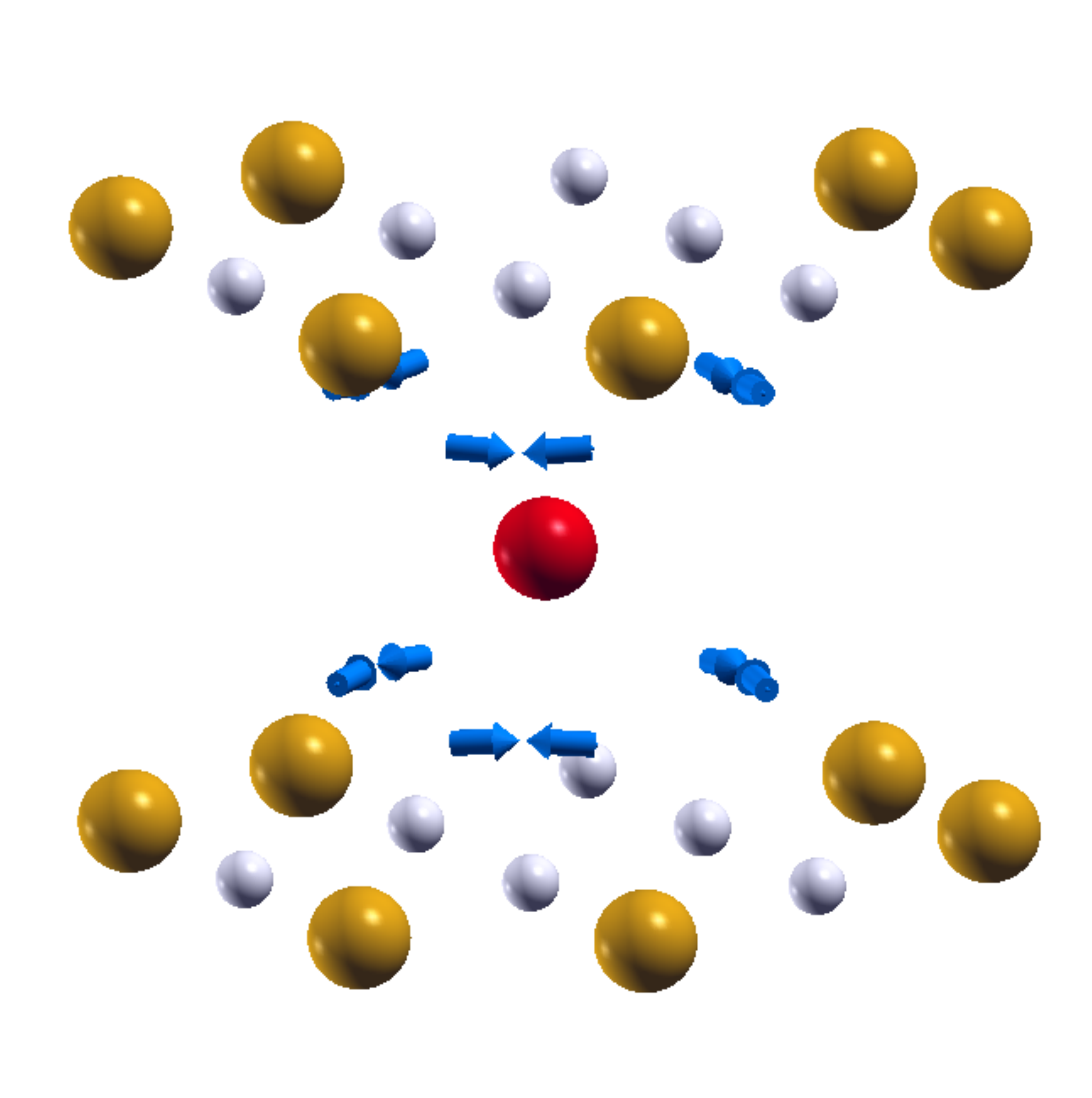}
\\
$|\vec{D}_{02}| = 0.21 meV$ \;\;\;\;\;$|\vec{D}_{03}| = 0.13 meV$
\;\;\;\;\;$|\vec{D}_{05}| = 0.08 meV$ \\
$|D^z_{02}| = 0.09 meV$ \;\;\;\;\;$|D^z_{03}| = 0.09 meV$
\;\;\;\;\;$|D^z_{05}| = 0.01 meV$ \\\; \; \;(c)
\\
\includegraphics[width=0.15\textwidth,angle=0,clip]{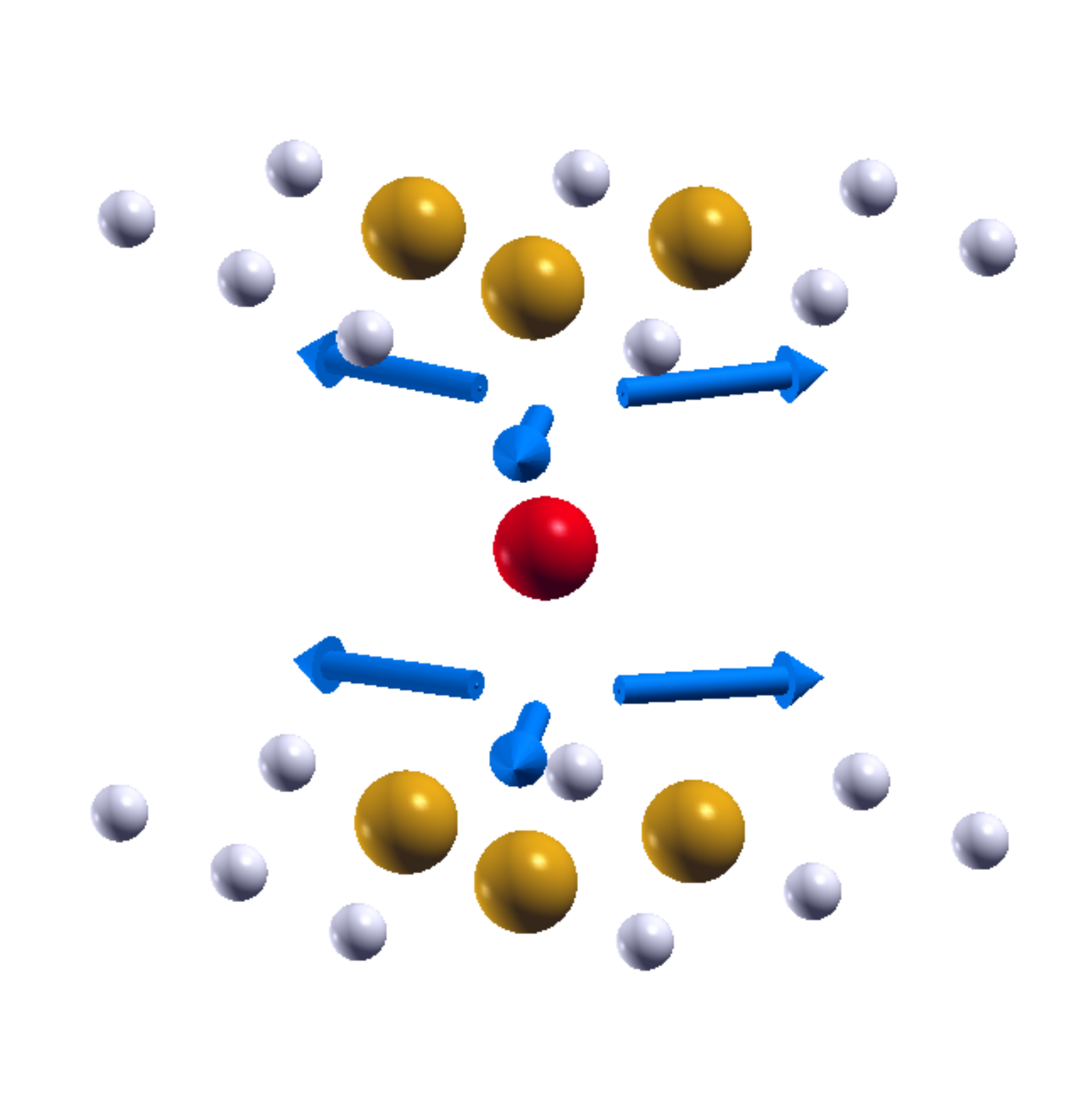}
\includegraphics[width=0.15\textwidth,angle=0,clip]{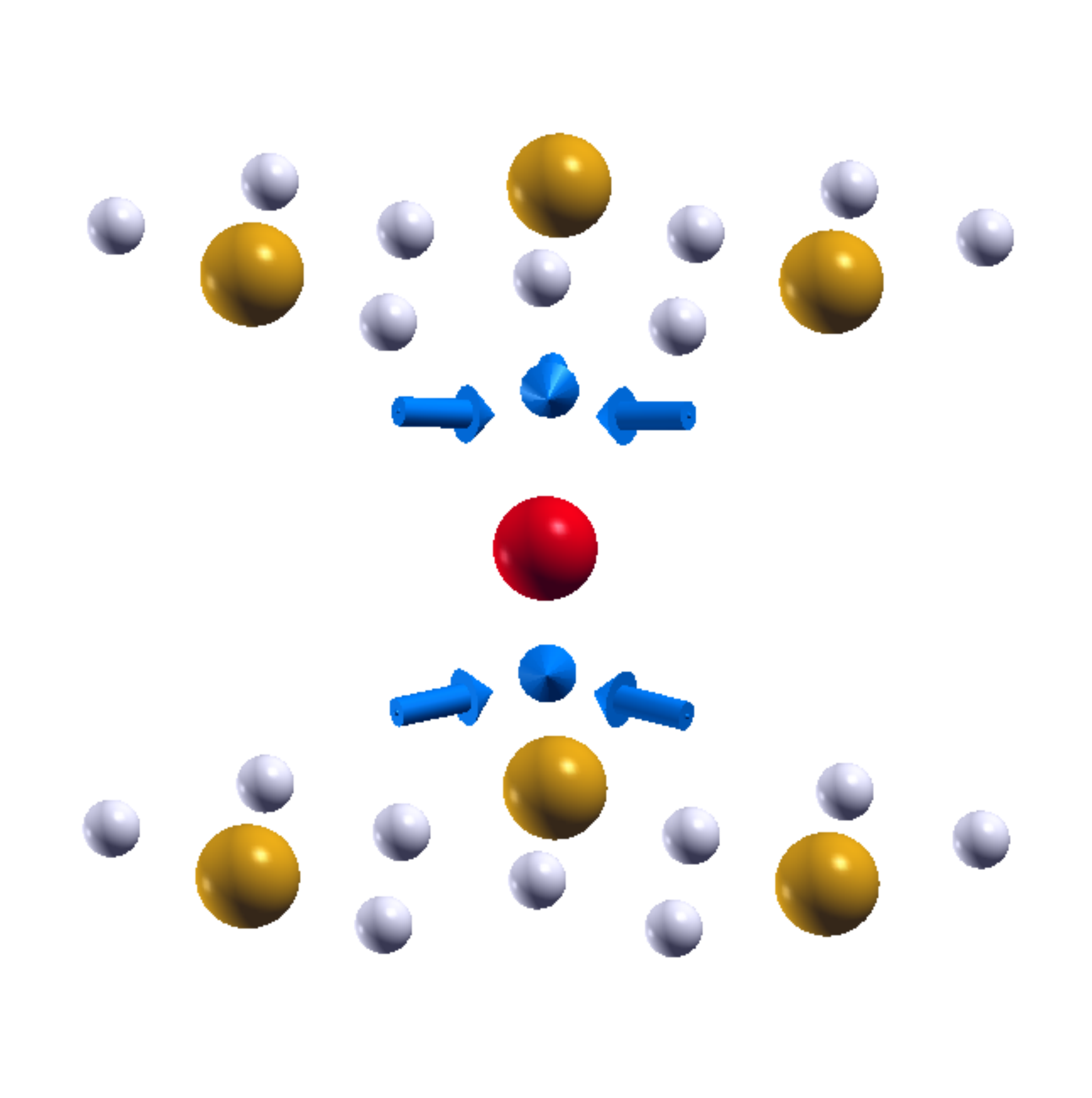}
\includegraphics[width=0.15\textwidth,angle=0,clip]{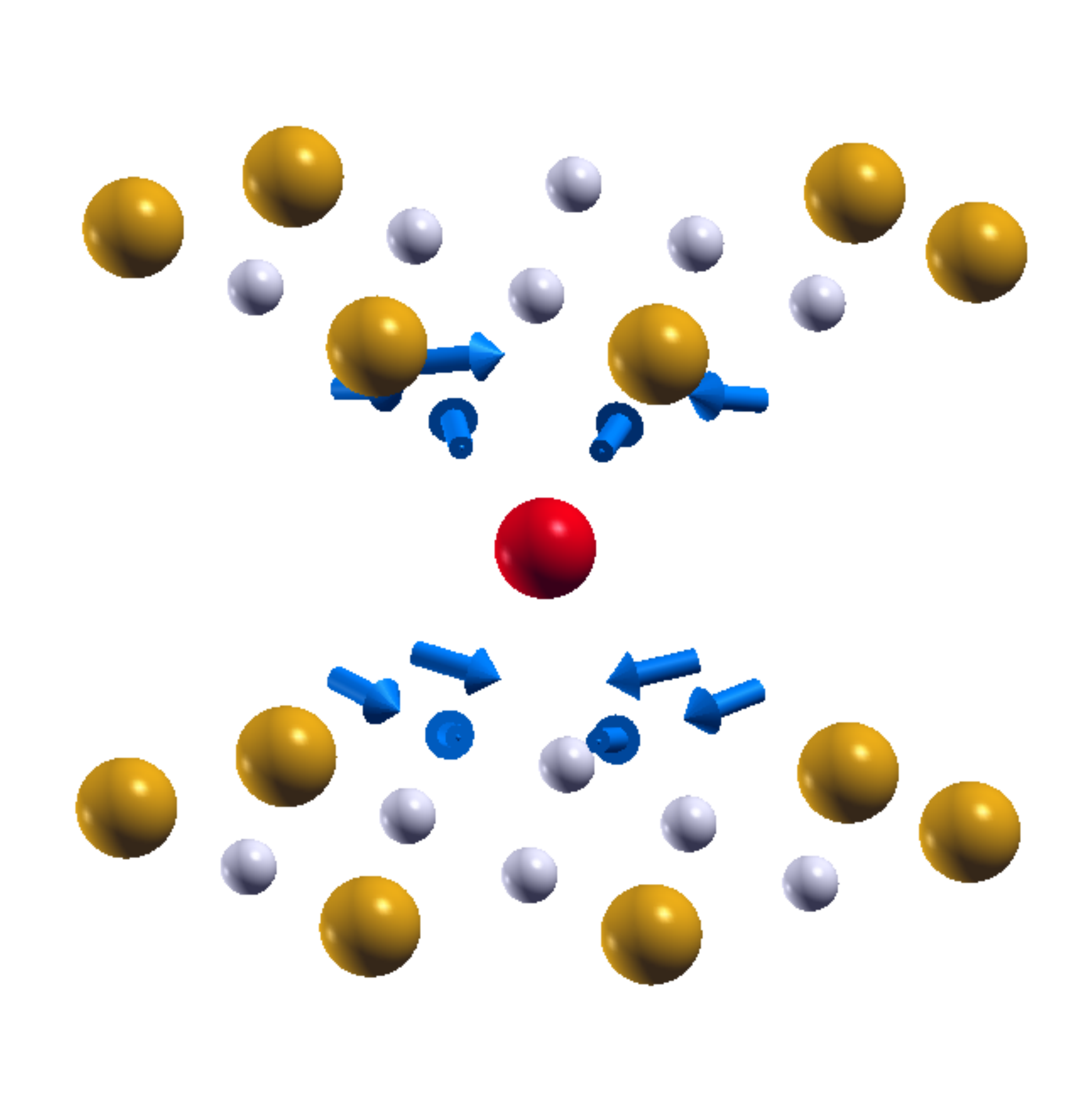}\\
$|\vec{D}_{02}| = 0.78 meV$ \;\;\;\;\;$|\vec{D}_{03}| = 0.59 meV$
\;\;\;\;\;$|\vec{D}_{05}| = 0.84 meV$  
$|D^z_{02}| = 0.02 meV$ \;\;\;\;\;$|D^z_{03}| = 0.07 meV$
\;\;\;\;\;$|D^z_{05}| = 0.16 meV$ \\
\; \; \;(d)
\caption{\label{fig:CMP_DM_Me} (a) Dzyaloshinskii-Moriya exchange coupling
  parameters shown schematically for all $M_{1/3}$NbS$_2$, between
  the first $\vec{D}_{01}$ and second $\vec{D}_{04}$ neighbors within
  the $M$ plane. DM interactions between the $M$ atoms corresponding to
  different neighboring planes: (b) for Cr$_{1/3}$NbS$_2$, (b)
  Mn$_{1/3}$NbS$_2$ and (c) Fe$_{1/3}$NbS$_2$.}  
\end{figure}

The calculated DM interactions for the compounds under discussion, namely 
the magnitudes and the directions of the DM vectors are shown in
Fig. \ref{fig:CMP_DM_Me}.
The intralayer DM interactions have the main component along the
lines connecting two atoms, with a small $z$ component as shown in
Fig. \ref{fig:CMP_DM_Me}(a). 
One can also see that the weakest interlayer interactions occur for
Cr$_{1/3}$NbS$_2$ while the largest for Fe$_{1/3}$NbS$_2$.  
The same behaviour is found for the $D^z$ components of the DM interactions
vector $\vec D$, 
although the $|D^z|/|\vec{D}|$  ratio decreases in the sequence from
Cr$_{1/3}$NbS$_2$ to Fe$_{1/3}$NbS$_2$. Note that the interlayer
components $D^z$ are responsible for the formation of a helimagnetic
structure along the $c$ axis in Cr$_{1/3}$NbS$_2$. 
Moreover, their magnitude decrease slowly with distance (see in
Fig. \ref{fig:CMP_DM_Me}), while the sign changes from shell to shell.
Therefore, the period of the HM structure is determined by the DM
interactions of Cr atoms which belong at least to several neighboring
shells where the DM interactions are non-negligible.
%

The calculated isotropic exchange parameters, DM interactions and the
anisotropy constant $K_1$ 
have been used as input for the MC simulations. As it was pointed out above, the FM
order has been obtained in the case of Cr$_{1/3}$NbS$_2$ and Mn$_{1/3}$NbS$_2$
compounds and the non-collinear AFM order in the case of
Fe$_{1/3}$NbS$_2$, when the DM interactions were not taken into
account. The calculated critical temperatures are listed in Table
\ref{TAB_TC} in comparison with the experimental results demonstrating
rather good agreement.  

\begin{figure}[h]
\includegraphics[width=0.3\textwidth,angle=0,clip]{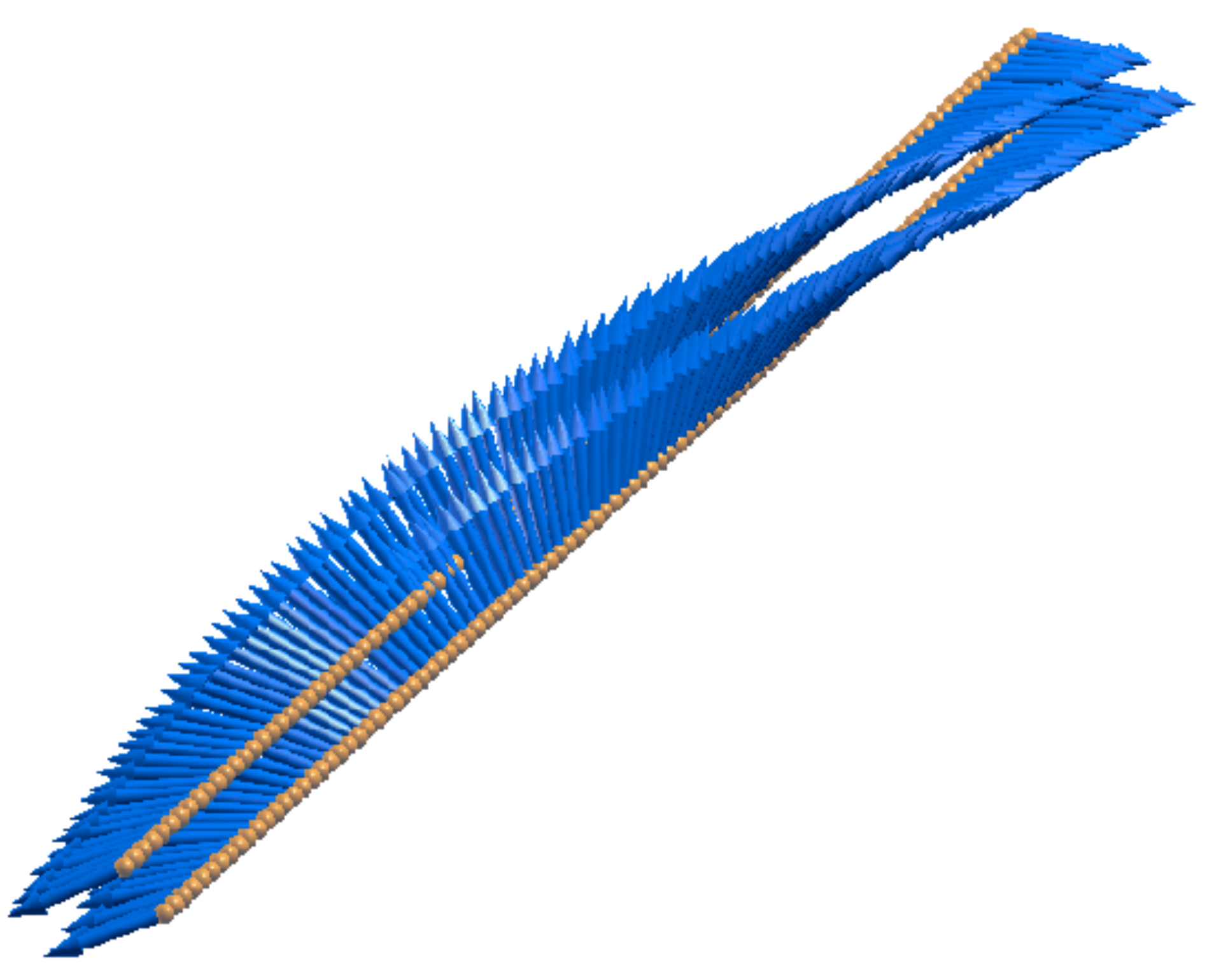}
\caption{\label{fig:MAG_CrNbS2} Helimagnetic structure along $c$ axis
  in Cr$_{1/3}$NbS$_2$ at $T = 1.0$~K. }  
\end{figure}

The interplay of the in-plane MCA and of the DM interactions
(namely, $D^z$ component) in Cr$_{1/3}$NbS$_2$ result in a long-period
helimagnetic structure represented by the in-plane Cr magnetic moments
twisted around the $c$ axis as it is shown by the snapshot in 
Fig. \ref{fig:MAG_CrNbS2} obtained from MC
simulations. Note that this result is
obtained  accounting only the first two neighbour shells. 
In this case, the period of the HM order is of $\sim 340a$ where $a$ is the lattice 
parameter, that significantly overestimates the HM period observed
experimentally $\sim 46$ nm, or $\sim 86a$, respectively, \cite{TKT+12}. 
The simple evaluation accounting for nearest neighbour interactions only
 gives a period similar to that obtained via MC simulations,
$2\pi a \frac{D^z_{02}}{J_{02}} = 375a$. To calculate the magnetic
structure more accurately, one has to take into account the exchange
interactions within the sphere of $~4a$ because of the oscillating behaviour
of the $D^z$ component of the DM interactions shown in  Fig. \ref{fig:CMP_JXC}(b)). 
A simple estimate leads to a HM period of
$~500a$, i.e. much longer than the experimental value.
A similar trend was also obtained for Mn$_{1/3}$NbS$_2$.
From these one can conclude that the DM interactions calculated in the
work are essentially underestimated. This behaviour correlates with
the results for the MCA that is also too low when compared to the
experiment. This can presumably be attributed to the approximation used for the
exchange-correlation functional. In the present work the LSDA was used,  
while use of the LSDA+U scheme could be more suitable to get better 
agreement with the experiment, as for example in the case of MCA
calculated for Fe$_{1/4}$TaS$_2$ system \cite{MCK+15}.
 
In the case of Mn$_{1/3}$NbS$_2$ no direct observations of the
helimagnetic structure have been reported, although, the transition 
to such a state has been observed \cite{KNK+09}.
This property can be associated with a rather long period of the
helimagnetic structure caused by the oscillating behaviour of the
$D^z$ component of the DM interactions (see Figs. \ref{fig:CMP_JXC}(b)), 
that can lead to some problems concerning its experimental observations.

\begin{table}

\begin{tabular}{ |c||c|c|c|c|c|  }
 \hline
               & MC & RDLM & EXPT \\
 \hline
  Cr$_{1/3}$NbS$_2$  &  115(FM)  & 310(FM) & 127(FM) \cite{MKK+83}  \\
  Mn$_{1/3}$NbS$_2$  &  80(FM) & 170(FM) & 65(FM) \\
  Fe$_{1/3}$NbS$_2$  & 63(AFM) & - & 47(AFM)\cite{GBR+81}, ~44(AFM)\cite{YMT+04} \\
\hline
\end{tabular}
\caption{\label{TAB_TC} The  Curie and N\'eel temperatures (in K units)
  calculated for $M_{1/3}$NbS$_2$ compounds via MC simulations and RDLM
  selfconsistent calculations, in comparison with experimental results. }        
\end{table}

\begin{figure}[h]
\includegraphics[width=0.2\textwidth,angle=0,clip]{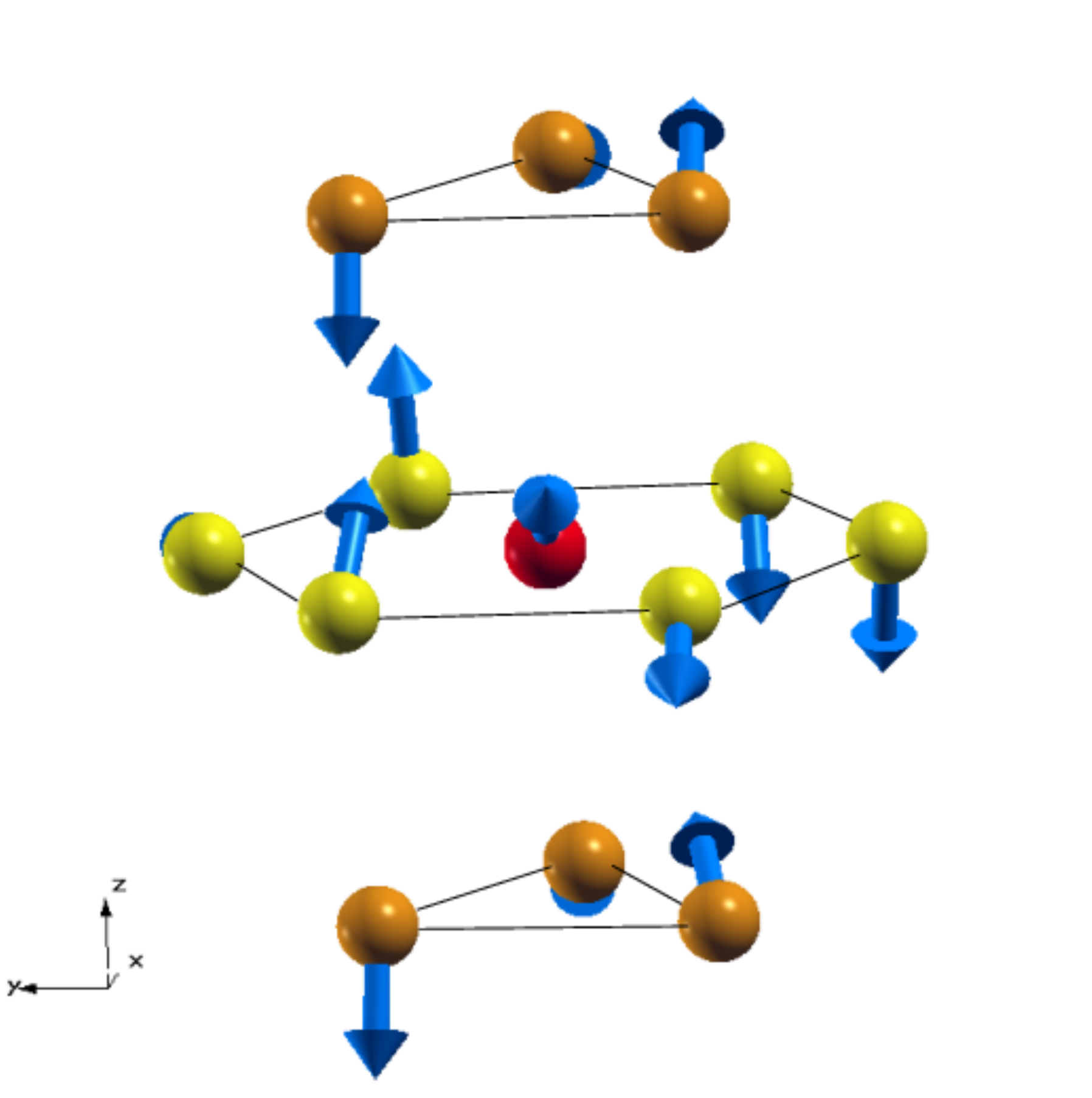}\;(a)
\includegraphics[width=0.2\textwidth,angle=0,clip]{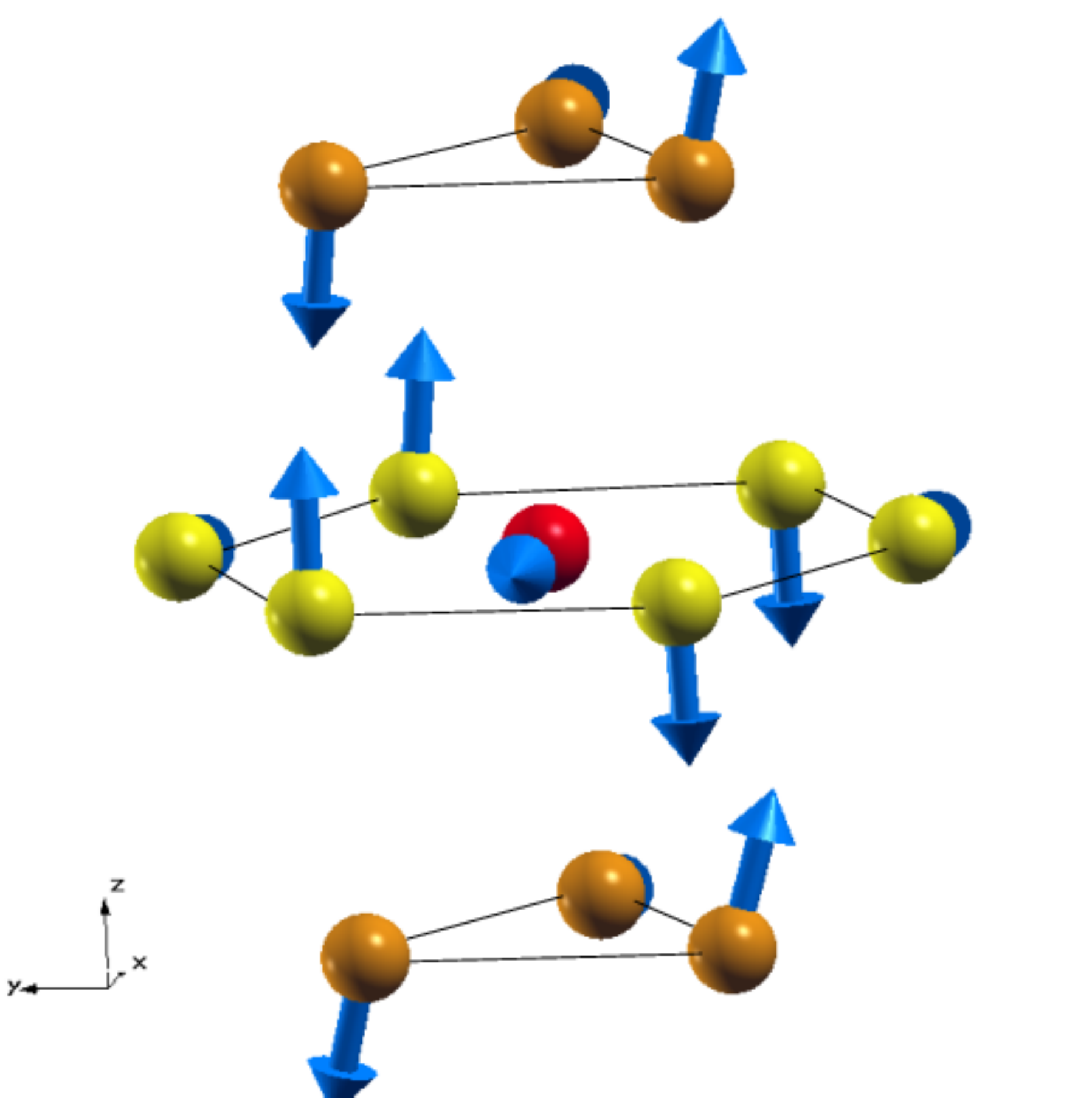}\;(b) \\
\includegraphics[width=0.2\textwidth,angle=0,clip]{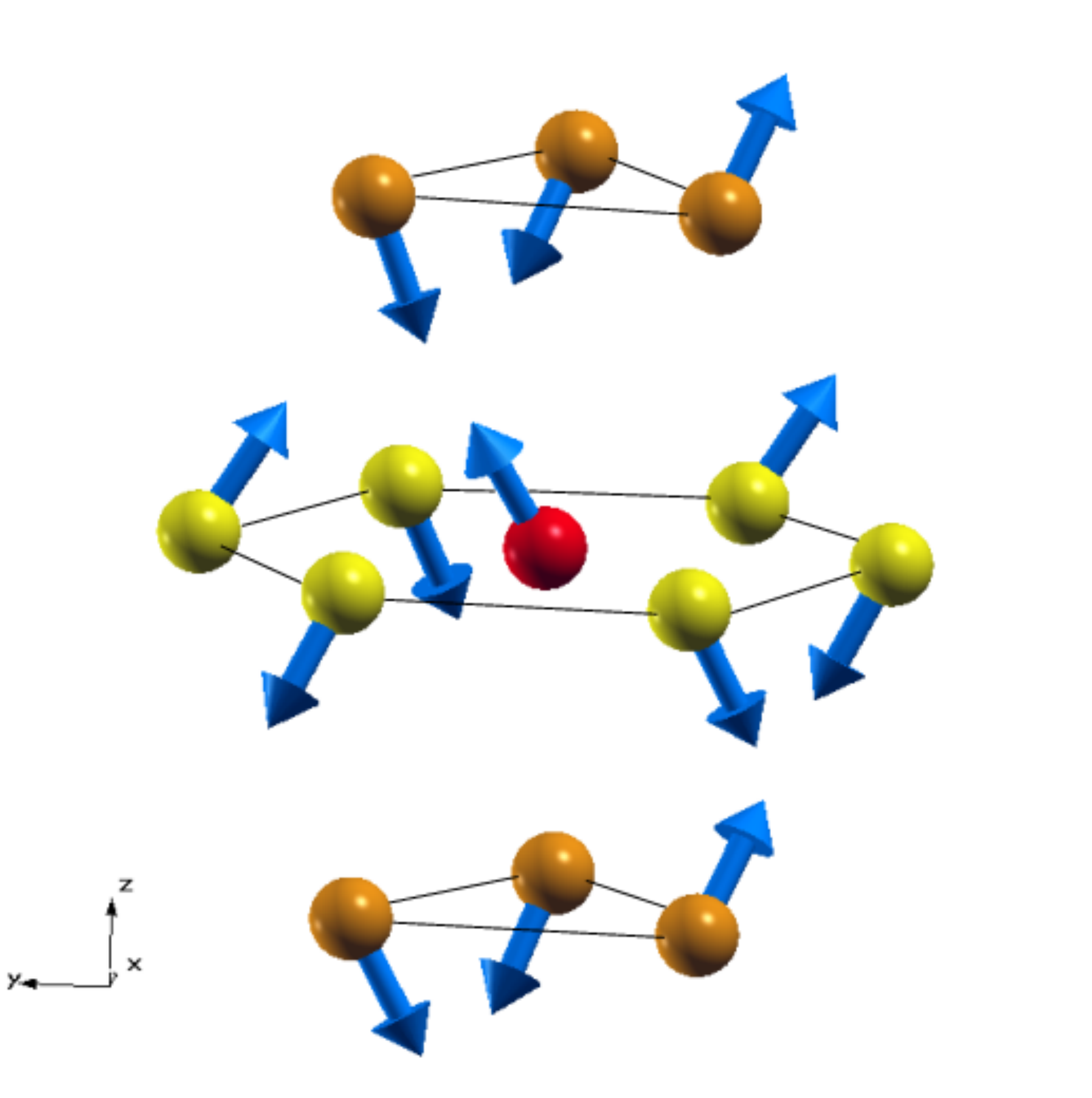}\;(c)
\includegraphics[width=0.19\textwidth,angle=0,clip]{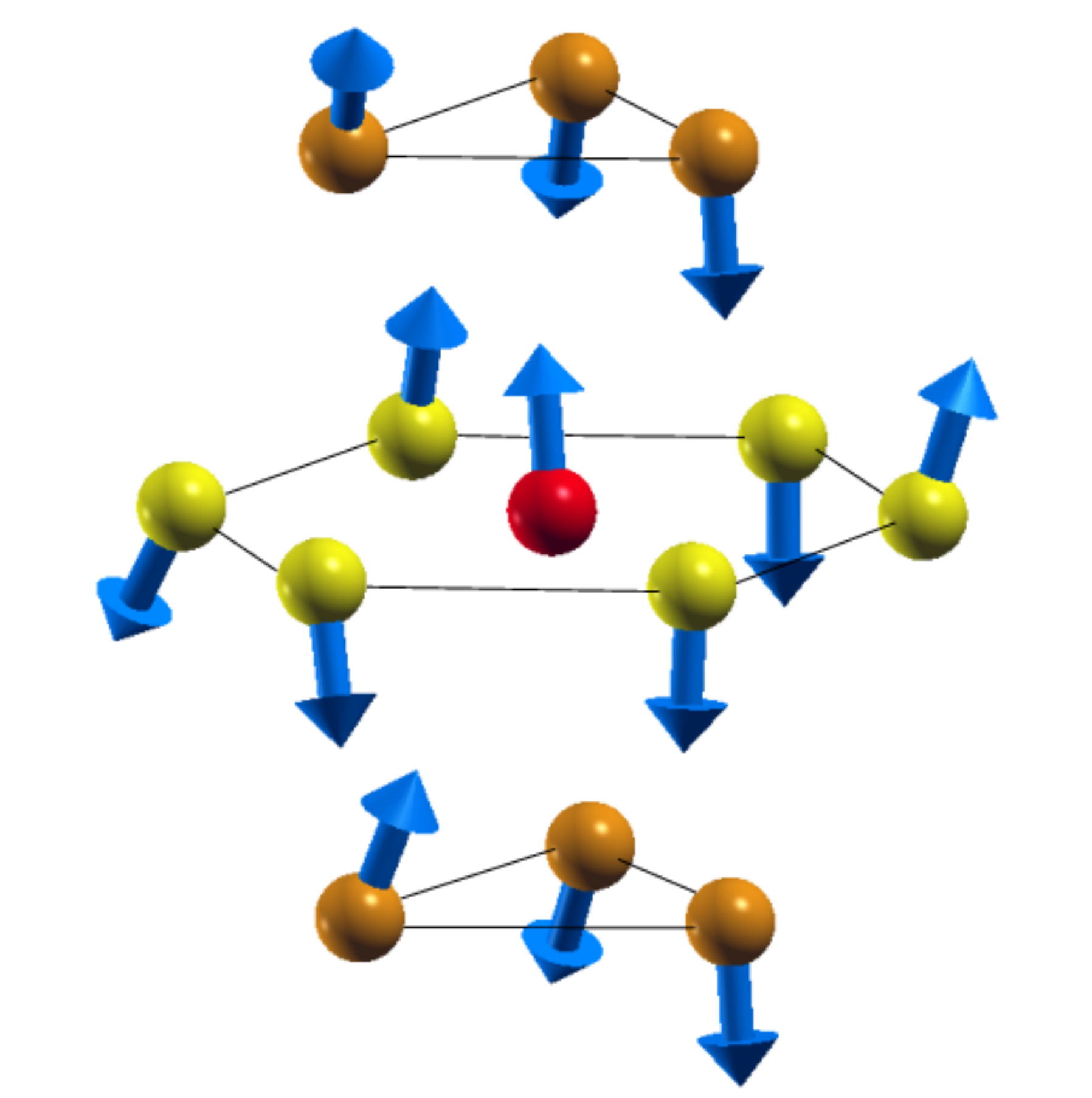}\;(d)
\caption{\label{fig:MAG_FeNbS2_2} Magnetic structure in
  Fe$_{1/3}$NbS$_2$ at $T = 1$~K: (a) DM interactions and MCA are not
  taken into account; (b) accounting for DM interactions, but without
  account of MCA; (c) accounting for DM interactions, with calculated
  MCA coefficient $K_1 = -0.78$~meV/Fe; (d) accounting for DM interactions, with 
  MCA coefficient $K_1 = -4.0$~meV/Fe, and temperature $T = 4$~K. }  
\end{figure}

For the case of Fe$_{1/3}$NbS$_2$, first of all, the isotropic exchange
interaction between the first and second neighbors are negative 
making their antiparallel alignment preferable. However, due to the
hexagonal crystal structure, a competition of the AFM interactions
between the neighboring atoms results in a frustrated non-collinear
magnetic structure in the system. Fig. \ref{fig:MAG_FeNbS2_2}(a)
represents the magnetic structure obtained within the MC simulations ($T
= 1$~K) neglecting DM interactions. 
Although the $D^z$ components of the DM vectors are much smaller 
than two others, $D^x$ and $D^y$,  the DM interactions result in the
formation of a chiral spin-spiral structure in Fe$_{1/3}$NbS$_2$ along $x$, $y$ as well as $z$
directions with a rotation angle of $90^o$ between neighboring atoms along
each direction (see Fig.  \ref{fig:MAG_FeNbS2_2}(b)).

It should be noted that in experiment \cite{LRI71,GBR+81}
a collinear ferrimagnetic structure is formed for Fe$_{1/3}$NbS$_2$, in
contradiction to the present calculations. A reason for this discrepancy may be attributed to
the magnetic anisotropy along the hexagonal $c$ axis, which has not been taken into
account within the calculations shown in Fig. \ref{fig:MAG_FeNbS2_2}(a) and (b).
According to the MC simulations based on the Heisenberg model, the calculated
$K_1$ parameter for the magnetic anisotropy represented in Table \ref{TAB_MOM},
$K_1 = 0.79$~meV/(Fe atom), is still too small to force the system to a 
collinear state. However, imposing a stronger MCA with $K_1 = 4$~meV/(Fe
atom), the magnetic moments are almost perfectly aligned along the $c$ axis, as it
is shown in Fig. \ref{fig:MAG_FeNbS2_2}(c).
Note that the latter results were obtained for the temperature 4~K,
to follow the experimental conditions \cite{LRI71,GBR+81},
leading to a finite amplitude of thermal fluctuations.
The magnetic structure shown in Fig. \ref{fig:MAG_FeNbS2_2}(c) 
with the so-called ordering of the third kind \cite{And50}, is in full
agreement with experiment. This implies an antiferromagnetic
ordering of each Fe atom with its 12 neighbors such that four of them
have their magnetic moments in the same direction and the other 8 atoms in the
opposite direction, as a result of the competition of their AFM
exchange coupling and uniaxial MCA along the $c$ axis.
Although, no experimental results on the MCA are available, 
the present results allow us to conclude 
about the underestimated MCA in our calculations, as it was already
shown for Cr$_{1/3}$NbS$_2$. Despite this, one can see
rather good agreement between theory and experiment, that gives
access to the understanding of the origin of magnetic structures and
magnetic properties of the systems under consideration.

\section{Summary}

In summary, first principles calculations for the exchange coupling, 
DM interaction and MCA parameters have been performed for Cr-, Mn-
and Fe-intercalated NbS$_2$. The FM interactions, finite values of the $D^z$
interaction terms and the in-plane MCA in Cr$_{1/3}$NbS$_2$ lead to a
helimagnetic structure along the hexagonal $c$ axis, in full agreement with
experiment. Similar properties have been obtained  also for
Mn$_{1/3}$NbS$_2$. No direct experimental observation of the HM
structure in Mn$_{1/3}$NbS$_2$ has been done so far, that can be
attributed to its long period due to oscillating behaviour of the DM
interactions with a distance. In the 
case of the Fe$_{1/3}$NbS$_2$ compound, a frustrated 
non-collinear magnetic structure has been found in the present
calculations when the MCA is neglected. However, a strong MCA along the
hexagonal $c$ axis leads to the magnetic structure referred to as an
ordering of the third kind, being a consequence of a modification of
the frustrated non-collinear magnetic structure in the presence of
strong MCA.

\acknowledgments

 Financial support by the    Deutsche
  Forschungsgemeinschaft (DFG) via the priority programs SPP 1415
is acknowledged.


%

\end{document}